\documentclass[
 aps,
 amsmath,amssymb,
 reprint,
]{revtex4-2}

\usepackage{graphicx}
\usepackage{dcolumn}
\usepackage{bm}

\usepackage[utf8]{inputenc}
\usepackage[T1]{fontenc}
\usepackage{mathptmx}
\usepackage{etoolbox}

\makeatletter
\def\@email#1#2{
 \endgroup
 \patchcmd{\titleblock@produce}
  {\frontmatter@RRAPformat}
  {\frontmatter@RRAPformat{\produce@RRAP{*#1\href{mailto:#2}{#2}}}\frontmatter@RRAPformat}
  {}{}
}
\makeatother
\begin{document}

\preprint{AIP/123-QED}

\title[Conductance scaling in nanopore arrays: \\disentangling roles of pore arrangement and reservoir geometry]{Conductance scaling in nanopore arrays: \\disentangling roles of pore arrangement and reservoir geometry}

\author{Jacob Bair}
\affiliation{Department of Physics, Brown University, Providence, Rhode Island 02912, United States}
\author{Zachery Gottshall}
\affiliation{Department of Physics, Brown University, Providence, Rhode Island 02912, United States}
\author{Matthias Kuehne}
\email{kuehne@brown.edu}
\affiliation{Department of Physics, Brown University, Providence, Rhode Island 02912, United States}

\date{\today}

\begin{abstract}
The scaling of electrical conductance with the number of nanopores, $N$, in an array is of fundamental importance for applications in membrane separation and ion transport across biological membranes, and has been the subject of longstanding disagreement in the field. Here we resolve this disagreement for the case of pores with negligible surface charge by showing that two previously conflicting predictions for the scaling of conductance with $N$ are each correct in their own limiting reservoir geometry: cuboidal reservoirs commensurate with the interpore spacing, and pore arrays connected to infinitely sized reservoirs. Combining finite element simulations with theory, we show how the conductance scaling transitions between these two limits as reservoir size increases from the commensurate cuboidal geometry, for which the conductance scales linearly with $N$, towards an infinite hemisphere, for which the scaling of the conductance is sublinear in $N$ for closely spaced pores. For the case of infinite reservoirs, we show that a simple analytical model from electrical contact theory is mathematically equivalent to an existing point-charge capacitance model for nanopore arrays, and validate its applicability to arbitrary pore arrangements against finite element and method-of-moments calculations. Using this model, we derive a characteristic coupling length, $\ell_\mathrm{c,95}$, the interpore spacing at which the array conductance reaches 95\% of that of the same number of independent pores. $\ell_\mathrm{c,95}$ scales linearly with pore diameter $d$, and grows with $N$ as $\sim\log N$ for linear arrays and $\sim\sqrt{N}$ for square arrays, so that pore coupling persists to surprisingly large separations in large arrays. Our work provides a general framework for predicting nanopore array conductance as a function of pore arrangement and reservoir geometry, of importance to nanofluidics research and the targeted design of functional devices.
\end{abstract}

\maketitle

\section{Introduction}
Nanopores play important roles across various fields including cellular biology, membrane separation science, desalination, and energy harvesting~\cite{Schoch2008,Bocquet2010,Tagliazucchi2015,Aluru2023}. In particular, ion transport through nanopores has received sustained interest, as its detailed understanding underpins technologies in these fields~\cite{Chang2009, Sahu2019, Xiao2019, Wen2021, Wang2023}. The system typically studied in this context is one or multiple nanopores embedded in an insulating membrane and connected to electrolyte-filled fluid reservoirs on both ends of the membrane. The electrical resistance of this system can be described as an in-series connection of several resistances: the bulk resistances of the reservoirs, the bulk resistance of the nanopore(s), and the access resistance which captures the focusing of current streamlines into and out of the nanopore(s). When the pore length is much greater than the pore diameter, the electrical resistance is typically dominated by the bulk nanopore resistance, i.e., access effects become negligible. However, when the pore length is comparable to or smaller than the pore diameter, access effects become dominant. Pores in atomically thin 2D materials are typically access-dominated, as are, depending on aspect ratio, pores etched in thin membranes such as SiN, as well as biological pores in lipid bilayer membranes.

The single-sided access resistance of an individual circular nanopore with diameter $d$ has been established as~\cite{Hall1975}
\begin{equation}
\label{eq:hallResistance}
    R_\mathrm{a}=\frac{\rho_\mathrm{bulk}}{2d},
\end{equation}
where $\rho_\mathrm{bulk}$ is the bulk resistivity of the surrounding electrolyte solution. This expression assumes a semi-infinite electrolyte reservoir. For a nanopore of negligible thickness in contact with semi-infinite reservoirs on both sides of the membrane, the bulk resistance vanishes, so that the two access resistances are the sole contributions to the total resistance, and the conductance of a single pore becomes
\begin{equation}
\label{eq:hallConductance}
G_0=\left(2R_\mathrm{a}\right)^{-1}= \sigma_\mathrm{bulk}d.
\end{equation}
Here, $\sigma_\mathrm{bulk} = \rho_\mathrm{bulk}^{-1}$ is the bulk conductivity of the electrolyte solution. If instead one considers a finite-sized reservoir, the access resistance is reduced accordingly. Analytical expressions have been derived for specific finite-reservoir geometries, such as a cuboidal nanochannel connected to cuboidal reservoirs~\cite{Green2020}, or a cylindrical nanochannel connected to cylindrical reservoirs~\cite{Bair2026}. For a nanopore of arbitrary cross-section connected to a cuboidal reservoir, numerical expressions have also been reported for the access resistance~\cite{Sebastian2023}.

In many relevant scenarios one does not deal with a single pore but with multiple pores embedded in the same membrane. Since any realistic membrane has finite extent, pores cannot a priori be treated as infinitely far apart, in which case their conductances would simply add. The nature of nanopore coupling in such arrays is an area of increasing interest~\cite{Yang2022,Cao2025,Lin2026}. However, the scaling of conductance with the number of pores, $N$, in an array has been the subject of direct disagreement in the literature~\cite{Gadaleta2014,Lucas2020,Sebastian2023,Cao2024}. In particular, treatments of pore arrays surrounded by infinitely sized reservoirs predict sublinear conductance scaling with $N$ that depends on the pore arrangement in the array~\cite{Gadaleta2014}, while treatments based on commensurate, finite-sized reservoirs predict linear scaling with $N$ via an equivalent parallel-circuit picture~\cite{Sebastian2023}. These two results have been presented as competing descriptions of the same physical system, yet, as we show below, both are in fact correct --- each applying to a distinct limiting reservoir geometry rather than representing a general disagreement about the underlying physics.

Here, we resolve the controversy of conductance scaling in arrays of nanopores with negligible surface charge by showing that these two apparently conflicting predictions correspond to two limiting cases of a more general reservoir-geometry dependence: (1) linear scaling obtained for pore arrays in commensurate cuboidal reservoirs~\cite{Sebastian2023}, and (2) sublinear scaling obtained for pore arrays in infinite reservoirs~\cite{Gadaleta2014}. We show explicitly how the conductance scaling transitions between these two limits as the reservoir size is increased from the commensurate cuboidal array towards an infinite hemispherical reservoir. We demonstrate this for previously discussed linear and square pore arrangements, and extend the treatment to close-packed arrays such as may be relevant, e.g., to condensed arrays of biological~\cite{Grage2011} or synthetic~\cite{Shen2015} nanopores. For the case of infinite reservoirs, we show that a simple analytical model borrowed from electrical contact theory~\cite{Greenwood1966} is mathematically equivalent to the point-charge capacitance model previously proposed for nanopore arrays~\cite{Gadaleta2014}, and we validate the generality of this model against both finite element calculations and independent numerical method of moments~\cite{Gibson2021} calculations. Using Greenwood's theory of electrical constriction resistance~\cite{Greenwood1966}, we derive a characteristic nanopore coupling length, $\ell_\mathrm{c,95}$ —-- the center-to-center pore spacing at which the conductance of a regular array reaches 95\% of the conductance of the same number of independent, uncoupled pores. This coupling length scales linearly with pore diameter and exhibits the same asymptotic $N$-dependence as the geometric scaling factor of the conductance itself, $\sim\log N$ for linear arrays and $\sim\sqrt{N}$ for square arrays, so that coupling effects persist to surprisingly large pore separations in arrays with many pores. Finally, we derive a new analytical model for the conductance of a single nanopore connected to finite cuboidal reservoirs, assuming a radial inverse square root density within the pore, and validate it against finite element calculations. This comparison underscores that the commonly used assumption of a radially uniform current density, while adequate for long channels, is not appropriate for pores of vanishing length and leads to significant overestimation of the access resistance in this regime and therefore an underestimation of the conductance. Together, these results provide a unified and general framework for predicting the electrical conductance of nanopore arrays as a function of pore arrangement and reservoir geometry, of central importance to nanofluidics research.

\begin{figure*}[ht]
\includegraphics[width=165.17mm]{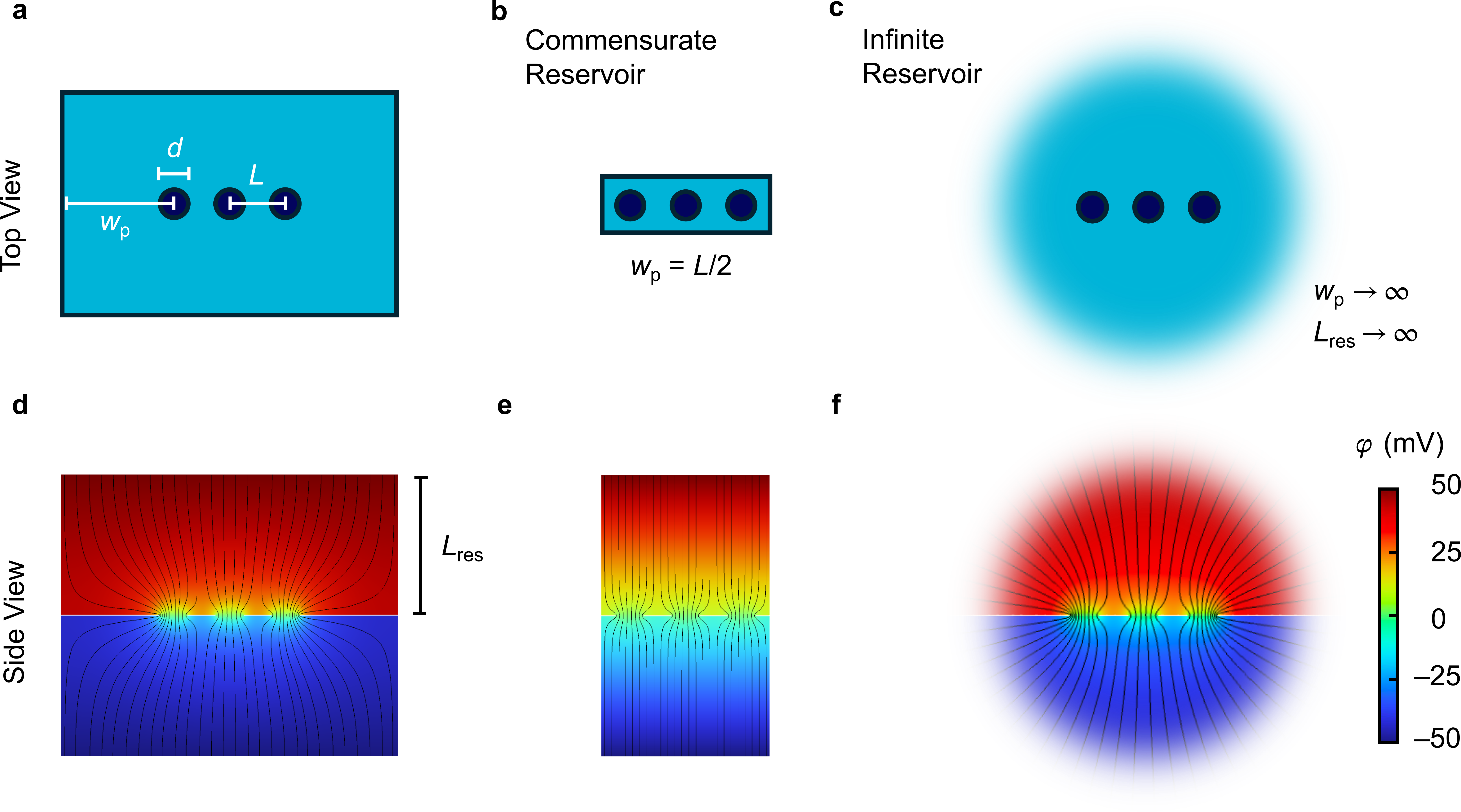}
\caption{\label{fig:01} Schematics of the various reservoir geometries located on each side of the nanopore array that were examined for linear arrays of identical nanopores of diameter, $d$, and negligible thickness separated by an interpore spacing, $L$. Top view cross sections in the $xy$-plane at $z = 0$ are shown for (a) finite sized cuboidal reservoirs of length $L_\mathrm{res}$ and a padding width, $w_\mathrm{p}$, that is larger than half of the distance between the pores ($w_\mathrm{p}>L/2$), (b) finite sized cuboidal reservoirs of length $L_\mathrm{res}$ and a padding width, $w_\mathrm{p}$, that is commensurate with half of the interpore spacing ($w_\mathrm{p}=L/2$), and (c) infinite hemispherical reservoirs ($w_\mathrm{p}, L_\mathrm{res} \rightarrow \infty$). The corresponding side view cross sections in the $xz$-plane at $y = 0$ are shown below with current streamlines and equipotential false coloring for (d) the incommensurate finite cuboidal reservoirs, (e) the commensurate finite cuboidal reservoirs, and (f) the infinite hemispherical reservoirs.}
\end{figure*}

\section{Methodology}

\begin{figure*}[ht]
\includegraphics[width=165.17mm]{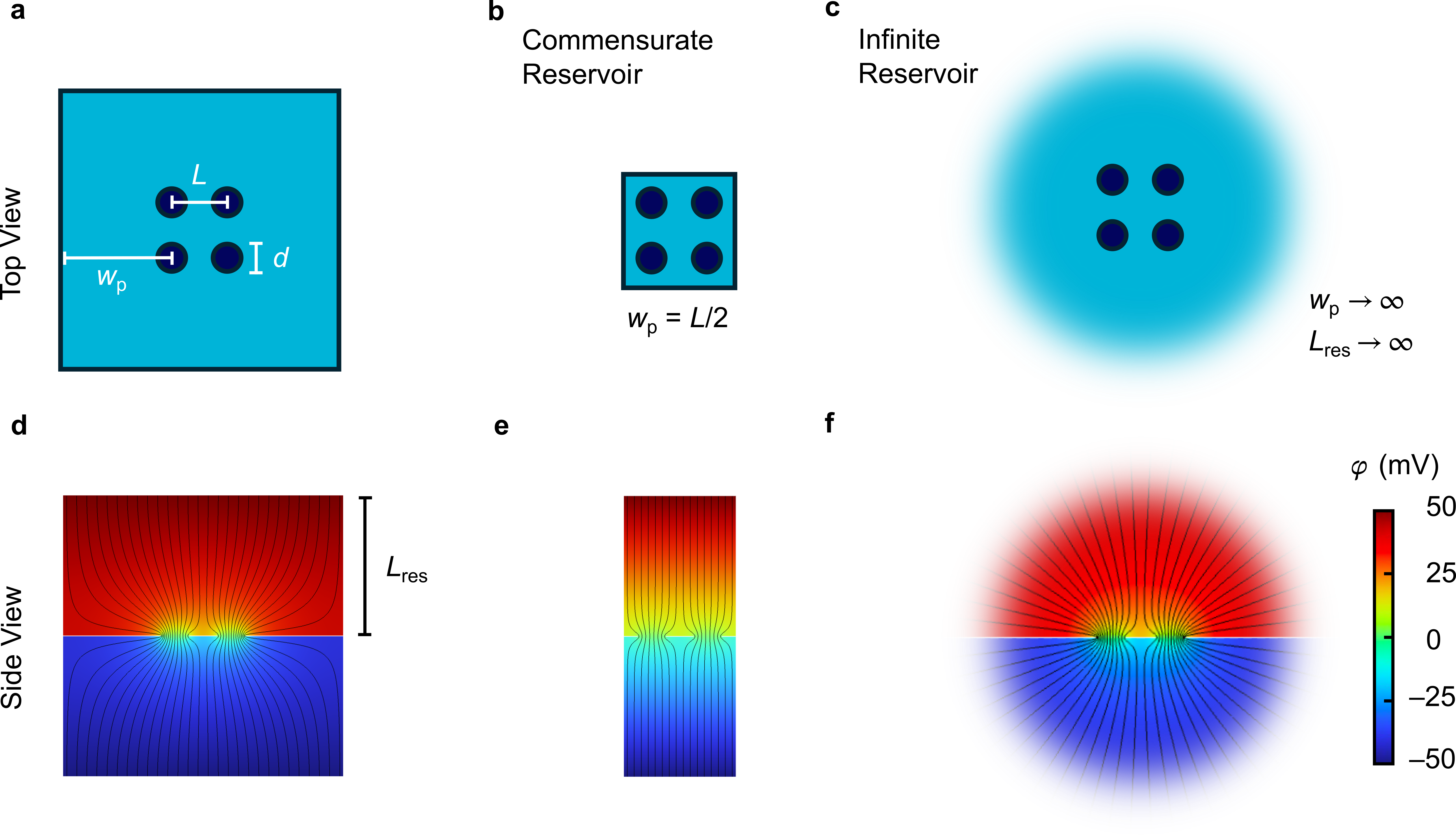}
\caption{\label{fig:02} Schematics of the various reservoir geometries located on each side of the nanopore array that were examined for square arrays of identical nanopores of diameter, $d$, and negligible thickness separated by an interpore spacing, $L$. Top view cross sections in the $xy$-plane at $z = 0$ are shown for (a) finite sized cuboidal reservoirs of length $L_\mathrm{res}$ and a padding width, $w_\mathrm{p}$, that is larger than half of the distance between the pores ($w_\mathrm{p}>L/2$), (b) finite sized cuboidal reservoirs of length $L_\mathrm{res}$ and a padding width, $w_\mathrm{p}$, that is commensurate with half of the interpore spacing ($w_\mathrm{p}=L/2$), and (c) infinite hemispherical reservoirs ($w_\mathrm{p}, L_\mathrm{res} \rightarrow \infty$). The corresponding side view cross sections in the $xz$-plane at $y = L/2$ are shown below with current streamlines and equipotential false coloring for (d) the incommensurate finite cuboidal reservoirs, (e) the commensurate finite cuboidal reservoirs, and (f) the infinite hemispherical reservoirs.}
\end{figure*}

We consider two types of nanopore array geometries: linear and square. Linear array geometries consist of a single line of $N$ pores, while square array geometries consist of a grid of $N$ pores divided into $\sqrt{N}$ rows and $\sqrt{N}$ columns. We assume nanopores of equal diameter, $d$, and negligible thickness. We define the interpore spacing, $L$, to be the center-to-center distance of nearest neighbor nanopores. The nanopores lie in the $xy$-plane located at $z = 0$, and are connected to exterior reservoirs of equal dimensions on both sides of the membrane, as would be the case in a nanofluidics experiment. Unless stated otherwise, we consider similarly sized exterior reservoirs of length $L_\mathrm{res}$, and a width and height determined by the geometry of the nanopore array and the padding width, $w_\mathrm{p}$, which we define as the perpendicular distance between the line passing through the centers of the outermost row or column of nanopores in the array to the reservoir wall to which the line is parallel. A voltage difference $\Delta V$ is applied across the model, with the top side held at $\frac{\Delta V}{2}$ and the bottom side at $-\frac{\Delta V}{2}$. The reservoirs are filled with a symmetric, monovalent, binary electrolyte characterized by its bulk conductivity
\begin{equation}
\label{eq:conductivity}
\sigma_\mathrm{bulk} = \frac{c_02e^2D}{k_\mathrm{B}T},
\end{equation}
where $k_\mathrm{B}$ is Boltzmann's constant, $T$ is the temperature of the system, $c_0$ is the salt concentration, $e$ is the elementary charge, and $D$ is the diffusion coefficient (assumed identical for the two ions). The outer walls of all of the compartments are assumed to be rigid and impermeable.

We consider finite length reservoir geometries for both commensurate reservoirs, for which the padding width is equal to half of the interpore spacing, $w_\mathrm{p} = L/2$, so that the geometry can be considered to be composed of $N$ identical units, and incommensurate reservoirs, for which $w_\mathrm{p}>L/2$. We also consider the case of infinite reservoirs for which both $w_\mathrm{p}\rightarrow\infty$ and $L_\mathrm{res}\rightarrow\infty$. Fig.~\ref{fig:01} shows the various geometries that were investigated for a linear array of nanopores and Fig.~\ref{fig:02} shows the various geometries that were considered for a square array of nanopores.

\subsection{Numerical Computation}
Finite element calculations were performed for the various geometries using COMSOL Multiphysics. We used a continuum-based approach and considered the case of negligible surface charge on the membrane. A voltage difference of $\Delta V = 0.1~\mathrm{V}$ was chosen for the calculations so that the top side of the system was held at 0.05~V and the plane containing the nanopores was held at 0~V due to the symmetry of the system and therefore the antisymmetry of the electric potential. Infinite reservoirs were approximated as hemispherical reservoirs of finite radius $R\gg d$, ensuring that the radius of the hemispheres was also much larger than the size of the nanopore array, $R\gg LN/2$ for linear arrays and $R\gg L\sqrt{N}/2$ for square arrays, for all values of $L$. For all calculations, the nanopore diameter was arbitrarily set to be $d = 4~\mathrm{nm}$ without loss of generality. Since we present our results only in terms of dimensionless ratios of the characteristic length scales of the system, the actual diameter used for the calculations is unimportant. However, since we use a continuum-based approach to model the system, it is important to note that the results can only be expected to hold for pore diameters above roughly $2-4$ nm, as below this threshold the granularity of the fluid, the finite size of the ions, and desolvation effects are expected to lead to the breakdown of the continuum description, although the exact limits of its validity remain a subject of debate~\cite{Faucher2019,Kavokine2021, Aluru2023}. Ion-ion correlations may also render the continuum description invalid at high electrolyte concentrations greater than about 0.1 M~\cite{Faucher2019}.

Since we considered the case of no surface charge on the membrane, the system is governed by Laplace's equation for the electric potential, $\nabla^2\varphi=0$. The walls of the system were set to be rigid and impermeable so that $\bm{n}\cdot\bm{\nabla}\varphi=0$ at the walls, where $\bm{n}$ is the normal vector to the wall. The current was then obtained as
\begin{equation}
\label{eq:current}
I= \sigma_\mathrm{bulk}\int_S{-\bm{\nabla}\varphi \cdot d\bm{S}}
\end{equation}
where $S$ is the cross-sectional area of the system in the $xy$-plane for the case of finite cuboidal reservoirs and a hemispherical surface for the case of infinite reservoirs approximated by large hemispherical reservoirs. The total conductance of the system was then calculated as $G = I/ \Delta V$. Fig.~\ref{fig:01}(d), (e), and (f) show cross sections of different geometries considered for linear arrays of pores in the $xz$-plane with equipotential false coloring and current streamlines obtained from the finite element calculations. Fig.~\ref{fig:02}(d), (e), and (f) show cross sections of different geometries considered for square arrays of pores in the $xz$-plane with equipotential false coloring and current streamlines obtained from the finite element calculations. We also note here that the no flux condition of the electric field lines, corresponding to rigid and impermeable walls, is only enforced in this model at the outermost walls of the system. However, there will also naturally be no flux of the electric field lines across planes of symmetry~\cite{Green2020}. Therefore, for the case where the padding width equals half of the interpore spacing, $w_\mathrm{p} = L/2$, enforcing a no flux condition at the outermost walls of the system ensures that the same boundary conditions are being applied periodically throughout the system and is therefore equivalent to establishing a periodic repetition of a single pore unit cell, which is a description of the system that will later become important when discussing the scaling relationships that are observed for the different pore arrangements and reservoir geometries.

\section{Results}

We first present the scaling of the conductance with reservoir size for a single pore. We then consider the simplest case of a linear array of pores, corresponding to just two pores, and the simplest case of a square array of pores, corresponding to a 2x2 grid of pores. For these two cases, we first present results for infinite hemispherical reservoirs with varying interpore spacing to isolate the effects of the separation distance between pores and then present results for various finite size cuboidal reservoirs to isolate the effects of reservoir size on the total conductance. We then present results for $N$ pore arrays for the case of infinite hemispherical reservoirs and finite cuboidal reservoirs to show how the conductance scales with $N$, which follows naturally from the scaling behavior of the smaller systems.

\subsection{A Single Pore}

\begin{figure}[t]
\includegraphics[width=83.2mm]{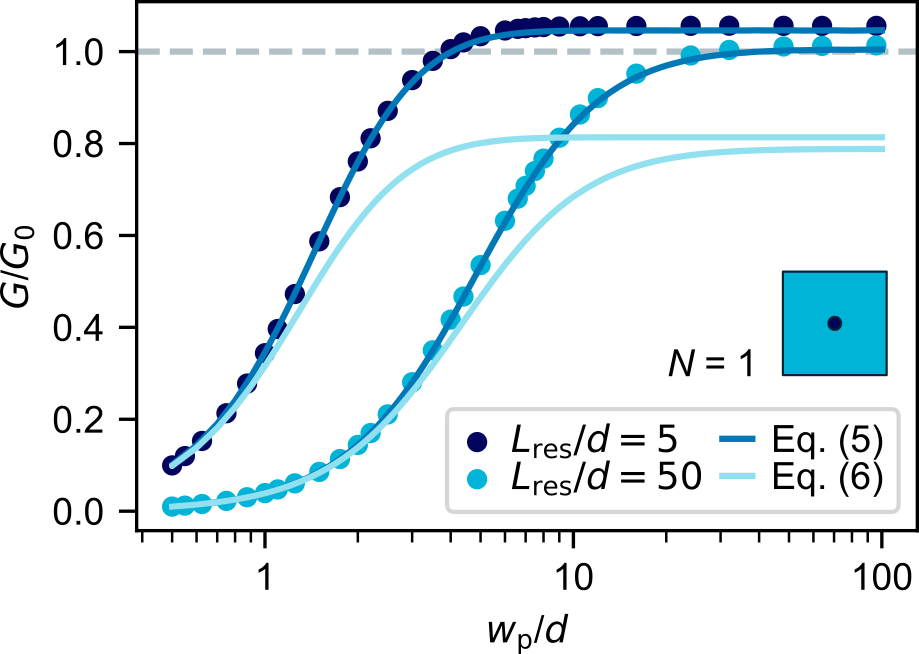}
\caption{\label{fig:03} The total conductance, $G$, normalized by $G_0$ for various ratios of the reservoir padding width, $w_\mathrm{p}$, to the diameter of the nanopore, $d$, for a single pore surrounded by cuboidal reservoirs of lengths $L_\mathrm{res}=5d$ and $L_\mathrm{res}=50d$. Dark blue solid curves indicate the predictions of Eq.~(\ref{eq:singalPoreModel}) and light blue solid curves show the predictions of Eq.~(\ref{eq:singalPoreModelConst}). A gray dashed line is located at $G/G_0 = 1$.}
\end{figure}

We first consider the case of a single nanopore. Since, for the case of infinite hemispherical reservoirs the conductance is given exactly by Eq.~(\ref{eq:hallConductance}), we consider only the case of finite size cuboidal reservoirs. Fig.~\ref{fig:03} shows the total conductance of the single pore system normalized by $G_0$ as given by Eq.~(\ref{eq:hallConductance}) for various ratios of the padding width, $w_\mathrm{p}$, to the pore diameter, $d$, ranging from 0.5 to 100. The data is plotted for two separate reservoir lengths, $L_\mathrm{res} = 5d$ and $L_\mathrm{res} = 50d$. For large values of $w_\mathrm{p}/d$, it can be seen that $G/G_0$ saturates at a value slightly higher than 1, as the bulk resistance of the system becomes negligible and the system becomes dominated by the contributions of the access resistance on each side of the pore. The access resistance on each side of the pore is lower than the value given by Eq.~(\ref{eq:hallResistance}) due to the finite length of the reservoirs, leading to the higher conductance. It can be seen that for the larger reservoir length, $L_\mathrm{res} = 50d$, the value at which this saturation occurs is closer to 1, as the increasing length of the reservoir brings the system closer to that of semi-infinite reservoirs. For small values of $w_\mathrm{p}/d$, $G/G_0\rightarrow d/2L_\mathrm{res}$, which is consistent with the bulk conductance of a single channel of cross-sectional area $d^2$ and length $2L_\mathrm{res}$, which is expected since for this case the access resistance goes towards zero and the bulk reservoir resistance, which is now nonzero due to the finite cross-sectional area of the system, controls the conductance.

In Fig.~\ref{fig:03} we also plot an analytical model for the conductance of a single nanopore surrounded by finite cuboidal reservoirs as solid curves. The model is derived assuming a radial inverse square root current density in the pore, which Rosenfeld and Timsit established as a very close approximation to the correct current distribution for a small constriction of vanishing length in electrical contact theory~\cite{Rosenfeld1981}, and which has since been validated for a cylindrical nanopore geometry by Bair et al.~\cite{Bair2026}. For comparison, we also derive the model assuming a radially constant current density, a common assumption for long nanochannels where the current has room to fully develop into a uniform profile~\cite{Sebastian2023}; as shown below, this assumption performs poorly for a pore of negligible thickness. For a radial inverse square root current density, the conductance is given by
\begin{equation}
\label{eq:singalPoreModel}
G = \sigma_\mathrm{bulk}\left[\frac{L_\mathrm{res}}{2w_\mathrm{p}^2}+\sum_{(m,n)\ne(0,0)} \frac{\sin{(\kappa_{mn}d/2)\tanh{(\kappa_{mn}L_\mathrm{res})}}}{w_\mathrm{p}^2d\mathcal{N}_m\mathcal{N}_n\kappa_{mn}^2}\right]^{-1}
\end{equation}
where $\kappa_{mn}=\frac{\pi}{2w_\mathrm{p}}\sqrt{m^2+n^2}$, and $\mathcal{N}_0=2$ and $\mathcal{N}_{i\ge1}=1$ with $i=m,n$.
For a constant current density the conductance becomes
\begin{equation}
\label{eq:singalPoreModelConst}
G = \sigma_\mathrm{bulk}\left[\frac{L_\mathrm{res}}{2w_\mathrm{p}^2}+\sum_{(m,n)\ne(0,0)} \frac{2J_1{(\kappa_{mn}d/2)\tanh{(\kappa_{mn}L_\mathrm{res})}}}{w_\mathrm{p}^2d\mathcal{N}_m\mathcal{N}_n\kappa_{mn}^2}\right]^{-1}
\end{equation}
where $J_1$ is the first order Bessel function. The derivation of these expressions is given in Appendix~A. As can be seen from Fig.~\ref{fig:03}, the radial inverse square root current density assumption provides much better agreement with the results of the finite element calculations than the assumption of a constant current density. As has been discussed previously~\cite{Bair2026}, the correct choice of the current density within the pore is essential in making accurate quantitative predictions for cases where the access resistance dominates, as, although the constant current density is able to predict a similar dependence on the dimensions of the reservoirs, the constant current density model significantly underestimates the value of the conductance for large reservoir widths.

\subsection{Two pores and Four Pores}

It has been pointed out previously that there can be no flux of the current streamlines through planes of symmetry~\cite{Sebastian2023}. This no flux condition is equivalent to adding rigid impermeable walls at these planes of symmetry. Therefore, for the case of multiple pores, we expect the scaling of the conductance to be dependent not only on the distance between the array and the outer walls of the reservoir, but also on the distance between the pores and the geometry of the nanopore array, which determine the locations of these planes of symmetry. To examine these different effects, we look at the simplest case of two different multipore array geometries: a linear array of two pores and a square array of four pores.

\subsubsection{Effects of Pore Spacing}

\begin{figure*}[t]
\includegraphics[width=165.17mm]{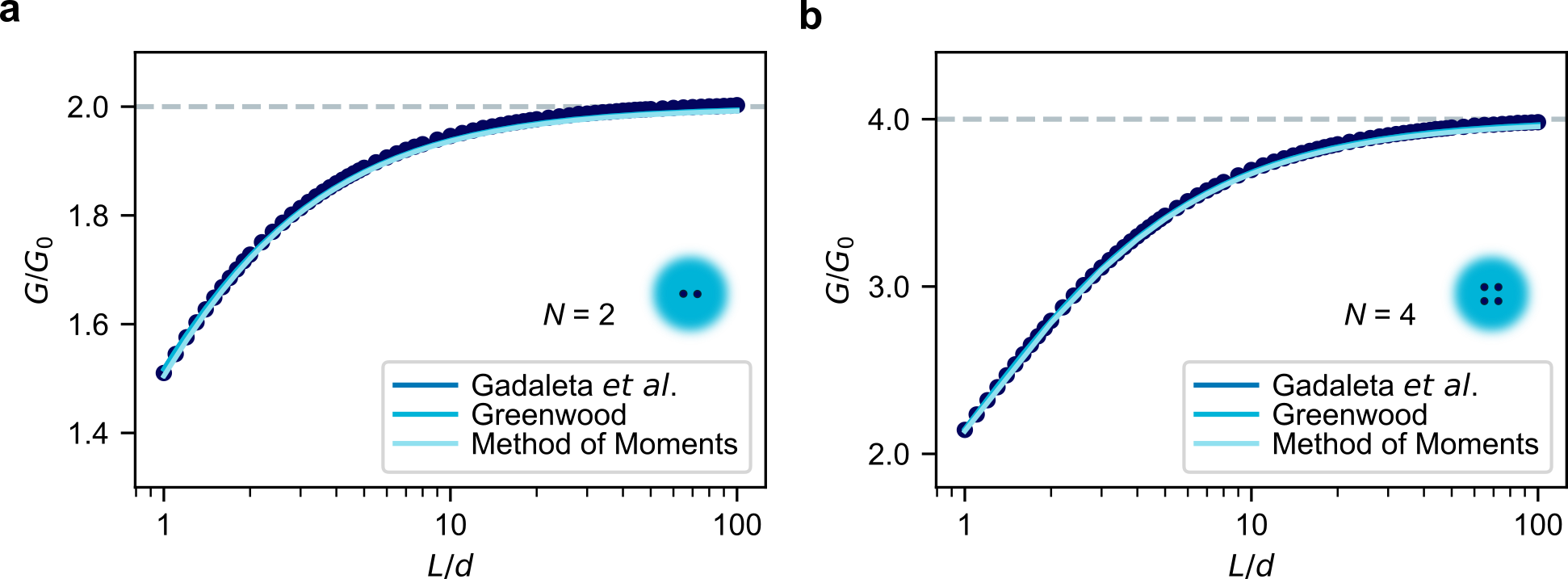}
\caption{\label{fig:04} The total conductance, $G$, normalized by $G_0$ for various ratios of the interpore spacing, $L$, to the diameter of the nanopores, $d$, for (a) two pores surrounded by large hemispherical reservoirs and (b) four pores in a square array surrounded by large hemispherical reservoirs. The predicted normalized conductances using Eq.~(\ref{eq:Gadaleta2Pore}) and Eq.~(\ref{eq:GadaletaNPore}) and the data obtained by Gadaleta et al.~\cite{Gadaleta2014}, Eq.~(\ref{eq:Greenwood}) as derived by Greenwood~\cite{Greenwood1966}, and the method of moments as described in Appendix~B are plotted alongside these results. Dashed lines are located at (a) $G/G_0 = 2$ and (b) $G/G_0 = 4$. The curves from Gadaleta et al. and Greenwood overlap.}
\end{figure*}

To isolate the effects of pore spacing, we plot the normalized conductance, $G/G_0$, as a function of the ratio of the interpore spacing to the pore diameter, $L/d$, for a linear array of two pores surrounded by infinite hemispherical reservoirs in Fig.~\ref{fig:04}(a) and for a square array of four pores surrounded by infinite hemispherical reservoirs in Fig.~\ref{fig:04}(b). If the symmetry planes created by including additional pores are far enough away from the positions of the pores, for each pore, similar to the case of a single pore in a finite sized compartment when both the length of the reservoir and the padding width become large, we expect the associated conductance to approach $G_0$, so that the total conductance scales linearly with $N$. In Fig.~\ref{fig:04}(a), as $L/d$ increases, the value of $G/G_0$ does indeed approach 2, and in Fig.~\ref{fig:04}(b) the value of $G/G_0$ approaches 4. For smaller values of $L/d$, the total conductance of the system decreases towards a value that is dependent on the specific geometry of the array. For large $N$ and closely spaced pores, the absolute minimum value that the conductance approaches should be $\sqrt{N}G_0$ as the geometry approaches that of a single pore with an effective diameter of $d_\mathrm{eff}=\sqrt{N}d$, which we see is even approximately true for the case of two and four pores.

We further compare these results to existing models for the conductance of nanopore array systems. Gadaleta et al.~\cite{Gadaleta2014} calculated the conductance for nanopore arrays surrounded by infinite electrolyte reservoirs by solving for the analogous problem of the capacitance of an array of equipotential disks. The capacitance for the analogous capacitance problem and the conductance for the nanopore arrays are related by a simple transformation (Eq.~(\ref{eq:proportionality}))~\cite{Hall1975}. Gadaleta et al.~\cite{Gadaleta2014} approximated the disks as point charges and used image charges to recursively calculate the effects of multiple pores on the total capacitance for large interpore spacing to diameter ratios, $L/d \gg 1$. This procedure, corrected from the form originally reported by Gadaleta et al.~\cite{Gadaleta2014} to account for a small error in the calculation, yields
\begin{equation}
\label{eq:Gadaleta2Pore}
    G = 2\sigma_\mathrm{bulk}d\left(1+\frac{d}{\pi L}\right)^{-1}
\end{equation}
for two pores and
\begin{equation}
\label{eq:GadaletaNPore}
    G = N\sigma_\mathrm{bulk}d\left(1+\gamma_N\frac{d}{\pi L}\right)^{-1}
\end{equation}
for $N$ pores, where $\gamma_N$ is a geometric factor that depends on the arrangement of the pores. For large $N$, $\gamma_N \sim \log{N}$ for a linear array of pores, and $\gamma_N \sim \sqrt{N}$ for a square array of pores, as reported by Gadaleta et al.~\cite{Gadaleta2014}. Based on these findings, the conductance for a linear array of pores should scale approximately with $N/\log{N}$ and the conductance for a square array of pores should scale approximately with $\sqrt{N}$. The predictions of this model are plotted for the case of two pores in Fig.~\ref{fig:04}(a) and the case of four pores in Fig.~\ref{fig:04}(b). In both cases, the predictions agree well with the finite element calculations.

Another model from electrical contact theory proposed by Greenwood~\cite{Greenwood1966} uses the analogy between current flow problems and electrostatic charge distribution problems and similarly models the contacts as points. The problem of current flow between two large conductors through small circular contacts and the problem of ion transport between two large electrolyte reservoirs through negligibly thin nanopores are analogous as long as surface charge and hydrodynamic effects may be ignored, as in the case that we consider, and so the results derived for electrical contact theory should also apply to such nanofluidic systems. Using this approach, the potential for each pore is calculated as the sum of the self interaction term and the potentials from the other contacts due to the spreading of the current under the condition that the potential of all of the contacts must be kept equal. By writing the produced heat in terms of the total current through the system and the approximated potential, the resistance of the system is calculated. The procedure gives 
\begin{equation}
\label{eq:Greenwood}
    G = N\sigma_\mathrm{bulk}d\left(1+\frac{d}{\pi N}\mathop{\sum\sum}\limits_{i\neq j}\frac{1}{L_{ij}}\right)^{-1}
\end{equation}
for the conductance of the system where both $i$ and $j$ index all of the pores and $L_{ij}$ is the spacing between pore $i$ and pore $j$. Greenwood points out that, since the actual current distribution in a conductor will minimize the heat produced, the resistance will be overestimated, and therefore the conductance calculated in this way should be an underestimate. The predictions of this model are plotted in Fig.~\ref{fig:04}(a) and Fig.~\ref{fig:04}(b). In both cases, the model agrees quite well with the finite element calculations. Further, we find that the curves from Greenwood's model and the model developed by Gadaleta et al.~\cite{Gadaleta2014} overlap. The overlap is not a coincidence. Since both models treat the disks as point charges, these two approximation methods are exactly equivalent, with one using the analogy between capacitance problems and conductance problems and the other using the analogy between current flow problems and electrostatics problems.

We also use the method of moments, commonly used to numerically solve electrostatics problems~\cite{Gibson2021}, to solve for the analogous problem of the capacitance of an array of equipotential disks from which we are able to once again find the conductance of our system through a simple transformation as has been previously demonstrated for single nanopores by Liu et al.~\cite{Liu2017}. The details of the calculation are described in Appendix~B. As can be seen in both Fig.~\ref{fig:04}(a) and Fig.~\ref{fig:04}(b), the conductance of the nanopore arrays obtained in this way agrees well with the conductance obtained from the finite element calculations.

\subsubsection{Effects of Reservoir Geometry}

\begin{figure*}[t]
\includegraphics[width=165.17mm]{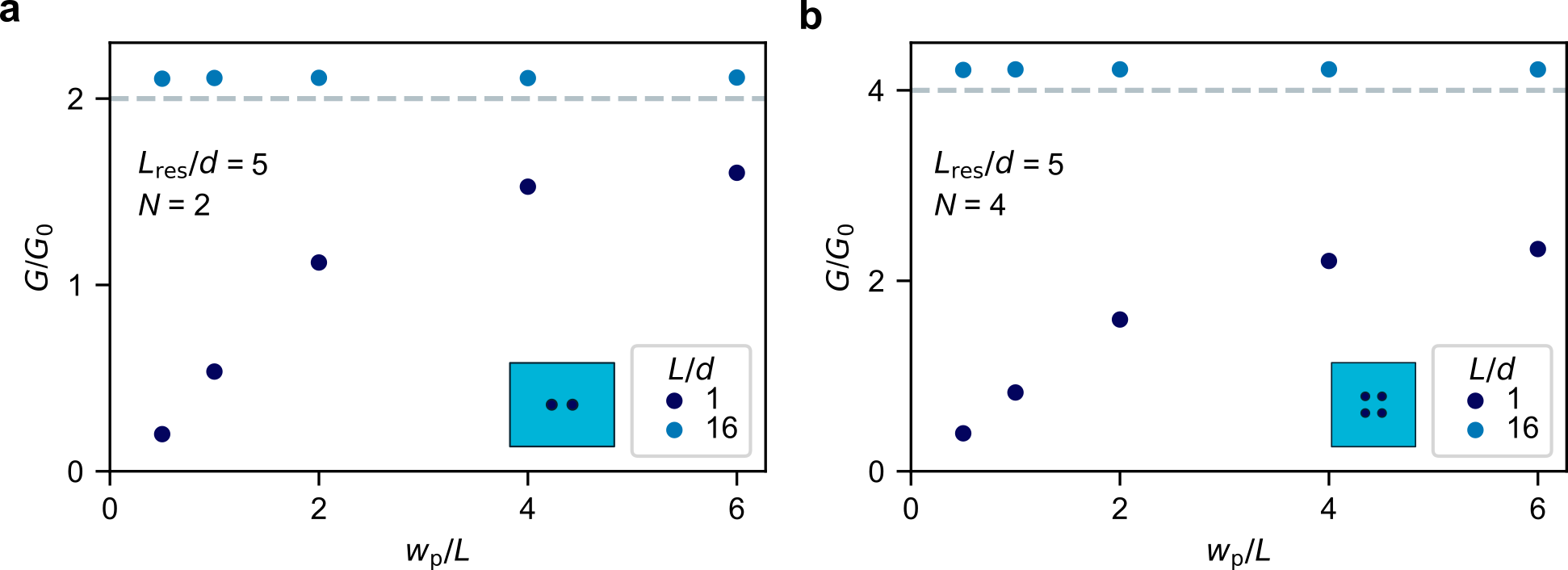}
\caption{\label{fig:05} The total conductance, $G$, normalized by $G_0$ for various ratios of the reservoir padding width, $w_\mathrm{p}$, to the interpore spacing, $L$, for large, $L/d=16$, and small, $L/d=1$, ratios of the interpore spacing to the nanopore diameter for (a) two pores surrounded by cuboidal reservoirs of length $L_\mathrm{res}=5d$ and (b) four pores in a square array surrounded by cuboidal reservoirs of length $L_\mathrm{res}=5d$. Dashed lines are located at (a) $G/G_0 = 2$ and (b) $G/G_0=4$.}
\end{figure*}

We now turn to exploring the effects of a finite size reservoir geometry on the scaling of the conductance for the two pore and four pore system. We plot the normalized conductance, $G/G_0$, as a function of the ratio between the reservoir padding width and the interpore spacing, $w_\mathrm{p}/L$, for the case of large, $L=16d$, and small, $L=d$, interpore spacing for the case of a linear array of 2 pores in Fig~\ref{fig:05}(a) and the case of a square array of 4 pores in Fig.~\ref{fig:05}(b). For all cases, the length of the reservoir is kept constant at $L_\mathrm{res}=5d$. We will discuss the effects of changing this length when we consider arrays with a larger number of pores. We note that, although we have chosen to look specifically at the limiting case here for small interpore spacing where $L = d$, the same general trends that are observed for this limiting case also hold for any small values of $L$ where $L/d\rightarrow1$.

For the case of large interpore spacing, $L/d=16$, we find the pores to essentially act independently as reflected in $G\approx2G_0$ and $G\approx4G_0$ for the linear array of two pores and the $2\times2$ square array of pores, respectively. Notably, this is independent of padding width. Note that the constant value of $G/G_0$ for these independent pores is slightly greater than $NG_0$ because the access resistance is less than that given by Eq.~(\ref{eq:hallResistance}) due to the finite length of the reservoirs. At the largest explored padding width ($w_\mathrm{p}/L=6$), reducing $L/d$ from 16 to 1, i.e., putting the pores extremely close together, reduces $G$ from approximately $2G_0$ to $\sqrt{2}G_0$ in the case of the linear array of two pores. This is consistent with the expectation that a dense array of pores approaches the conductance of a single pore with effective diameter $d_\mathrm{eff}=\sqrt{N}d$ mentioned previously. A similar behavior is observed for the $2\times2$ square array in Fig.~\ref{fig:05}(b) where at the largest padding width the conductance decreases from approximately $4G_0$ to $2G_0$. For the closely spaced pores, reducing the padding width has a pronounced effect on the normalized conductance: the conductance is observed to drop as $w_\mathrm{p}\rightarrow L/2$. This behavior is governed mostly by the bulk reservoir resistance which increases with decreasing $w_\mathrm{p}$ and dominates over the access resistance which decreases with decreasing $w_\mathrm{p}$. In the important limit of the commensurate case ($w_\mathrm{p}= L/2$), each of the two systems is an array of unit cells with unit cell conductance $G_0d/2L_\mathrm{res}$, so that $G/G_0$ approaches $2\times G_0d/2L_\mathrm{res}$ and $4\times G_0d/2L_\mathrm{res}$ in the case of the linear array of two pores and the $2\times2$ square array of pores, respectively.

\subsection{$N$ Pores, Infinite Reservoirs}

\begin{figure*}[p]
\includegraphics[width=165.17mm]{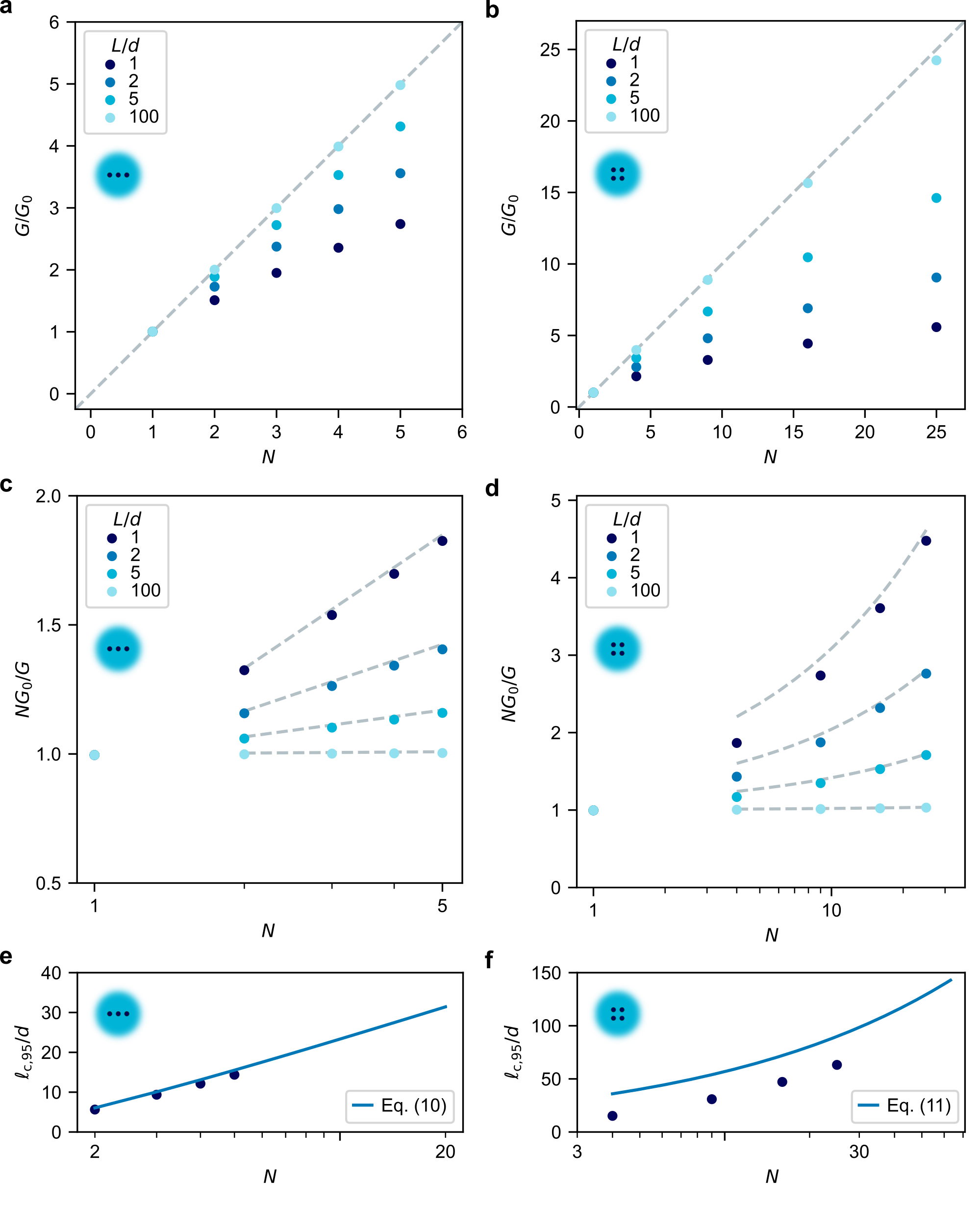}
\caption{\label{fig:06} The total conductance, $G$, normalized by $G_0$ for various ratios of the interpore spacing, $L$, to the nanopore diameter, $d$, for different values of the pore number, $N$, for (a) a linear array of pores surrounded by large hemispherical reservoirs and (b) a square array of pores surrounded by large hemispherical reservoirs. The number of pores, $N$, multiplied by $G_0$ and divided by the total conductance, $G$, is also plotted for different values of $N$ for $L/d = 1,\ 2,\ 5,\ \mathrm{and}\ 100$ for (c) a linear array of pores surrounded by large hemispherical reservoirs and (d) a square array of pores surrounded by large hemispherical reservoirs. Dashed lines are located at $G/G_0 = N$ in (a) and (b). In (c) and (d), dashed lines represent the predictions for the conductance scaling with the number of pores from Eq.~(\ref{eq:GadaletaNPore}) as originally reported by Gadaleta et al.~\cite{Gadaleta2014}. The ratios of the coupling length, $\ell_\mathrm{c,95}$, to the pore diameter, $d$, as extracted from the finite element method calculations are plotted for different pore numbers, $N$, for (e) a linear array and (f) a square array alongside predictions based Greenwood's theory (Eqs.~(\ref{eq:elllinear}) and (\ref{eq:ellsquare}), respectively), shown as solid blue curves.}
\end{figure*}

We return now to the case of infinite hemispherical reservoirs. The normalized conductance, $G/G_0$ is plotted for several values of $N$ for different ratios of the interpore spacing to the pore diameter, $L/d$, for both a linear array of pores in Fig.~\ref{fig:06}(a) and a square array of pores in Fig.~\ref{fig:06}(b). In both Fig.~\ref{fig:06}(a) and Fig.~\ref{fig:06}(b), dashed lines are included depicting the linear scaling relationship $G = NG_0$. For both the linear and square arrays, for large interpore spacings, the scaling approaches that of the dashed line. To resolve the sublinear scaling relationships better, we plot $N(G/G_0)^{-1}$ versus the number of pores, $N$, on a logarithmic scale for a linear array in Fig.~\ref{fig:06}(c) and for a square array in Fig.~\ref{fig:06}(d). In both cases, for a large interpore spacing, we observe a nearly horizontal line, indicating an approximately linear relationship in the scaling of the conductance with $N$ as expected. For a linear array, for small interpore spacing, the data points fall on a sloped line, indicating $\sim N/\log{N}$ scaling of the conductance as has been previously reported~\cite{Gadaleta2014}. Similarly, the data points for the square array form a curve indicative of $\sim \sqrt{N}$ scaling, in agreement with previous findings that report the scaling more precisely as $\gamma_N \approx N^{0.6}$ for cases where the large $N$ limit has not yet been reached~\cite{Gadaleta2014}. The predictions from Eq.~(\ref{eq:GadaletaNPore}) using the values of $\gamma_N$ originally reported by Gadaleta et al.~\cite{Gadaleta2014} are plotted alongside the data in Fig.~\ref{fig:06}(c) and Fig.~\ref{fig:06}(d) showing this agreement.

Using Greenwood's theory of electrical constriction resistance, we define a characteristic coupling length, $\ell_\mathrm{c,95}$, as the interpore spacing at which the conductance of a regular array reaches 95\% of the conductance of the same number of independent, uncoupled pores. We find that $\ell_\mathrm{c,95}$ scales linearly with pore diameter $d$, but also grows with pore number as $\sim\log N$ for a linear array (Fig.~\ref{fig:06}(e)) and $\sim\sqrt{N}$ for a square array (Fig.~\ref{fig:06}(f))---mirroring the same asymptotic $N$-dependence found for the geometric scaling factor of the conductance itself~\cite{Gadaleta2014}. For the linear array, an expression for $\ell_\mathrm{c,95}(N)= \ell_\mathrm{c,linear}(0.95,N)$ can be derived from Eqs.~(\ref{eq:hallConductance}) and (\ref{eq:Greenwood}) yielding (see Appendix~C):
\begin{equation}
\label{eq:elllinear}
    \ell_\mathrm{c,linear}(p,N)=\frac{2pd}{\pi(1-p)}\left(H_{N-1}+N^{-1}-1\right),
\end{equation}
where $H_n=\sum_{k=1}^nk^{-1}$ is the harmonic number. For large $N$, this expression asymptotically becomes $\ell_\mathrm{c,linear}(p,N)\approx\frac{2pd}{\pi(1-p)}\left(\log{N}+\gamma-1\right)$, where $\gamma\approx0.5772$ is the Euler constant, revealing the $\sim\log N$ scaling observed in Fig.~\ref{fig:06}(e). For the square array, an analytical expression for $\ell_\mathrm{c,95}(N)= \ell_\mathrm{c,square}(0.95,N)$ can be derived only in the large $N$ limit (see Appendix~C) where
\begin{equation}
\label{eq:ellsquare}
    \ell_\mathrm{c,square}(p,N)\approx\frac{4Kpd}{\pi(1-p)}\sqrt{N},
\end{equation}
with $K\approx0.74330$. This expression directly reveals the $\sim\sqrt{N}$ scaling observed in Fig.~\ref{fig:06}(f). Alongside these predictions, the values of the coupling lengths extracted from the finite element calculations are plotted in Fig.~\ref{fig:06}(e) and Fig.~\ref{fig:06}(f) as well. The coupling lengths extracted from the finite element calculations exhibit the expected scaling behaviors, although the values are lower than those predicted by Eq.~(\ref{eq:ellsquare}) for the square array. An exact match with Eq.~(\ref{eq:ellsquare}) is not expected, however, since the expression is only valid in the limit of large $N$.

We note here that our definition of the coupling length is completely arbitrary and is based on what a reasonable threshold may be for a drop in conductance that could be distinguished in an experiment from spurious effects related to tolerances and noise. In some contexts, it may be more desirable to define the coupling of the pores in an alternative way. For example, a more physically grounded definition of coupling could be obtained by considering the access region associated with the pores, the equipotential surface surrounding the pores within which a certain large percentage of the total resistance of the system is contained. For a single pore of negligible thickness surrounded by infinite reservoirs, the equipotential surfaces are oblate spheroids centered at the origin of the pore and defined by the equation
\begin{equation}
\label{eq:equipotentialSurface}
    \frac{4r^2}{d^2+4\mu^2}+\frac{z^2}{\mu^2} = 1,
\end{equation}
where $r$ and $z$ are the standard cylindrical coordinates, $d$ is the diameter of the pore, and $\mu$ is the length of the vertical semi-axis of a given equipotential spheroid as measured from the plane containing the pore~\cite{Holm1967, Slade2014}. The resistance within a single-reservoir half-space between the pore and the equipotential surface with semi-axis $\mu$ can be found as~\cite{Holm1967, Slade2014}
\begin{equation}
\label{eq:equipotentialExtent}
    R_{\mu} = \frac{\rho_\mathrm{bulk}}{\pi d}\arctan\left(\frac{2\mu}{d}\right).
\end{equation}
While the equipotential surfaces for a single pore have a well-defined spheroidal geometry, in general this is no longer the case when considering arrays of multiple pores. However, the access region of an array of pores may still be examined by considering quantities that are well defined for access regions of varying shape such as the volume contained between an equipotential surface and the plane of the pores or the maximum vertical extent of an equipotential surface from the plane of the pores. Here, we will look at the the maximum vertical extent of an equipotential surface from the plane of the pores, which we will call $z_{\mathrm{max},p}$, where $p$ defines the percentage of the access resistance contained within this region. For the case of a single pore, the maximum vertical extent of an equipotential surface is the same as the length of the vertical semi-axis of the equipotential spheroid so that $z_\mathrm{max} = \mu$ for each equipotential surface. The percentage of the total access resistance that the access region is defined to contain is arbitrary. Here we define the access region to be the region around the pore or pores that contains $95\%$ of the total resistance of the system. The vertical extent of this access region for a single pore can then be found analytically as
\begin{equation}
\label{eq:singlePoreExtent}
    z_{\mathrm{max},95} = \frac{d}{2}\tan\left(\frac{0.95\pi}{2}\right).
\end{equation}
For an array of pores, when the pores are closely spaced, as discussed previously, the array should behave as a single pore of effective diameter $d_\mathrm{eff} = \sqrt{N}d$, and, when the pores are far apart, the array should behave as a collection of independent pores. For these two limiting cases, $z_{\mathrm{max},95}$ can therefore be found using Eq.~\eqref{eq:singlePoreExtent} with the appropriately scaled diameter.

\begin{figure}[t]
\includegraphics[width=83.2mm]{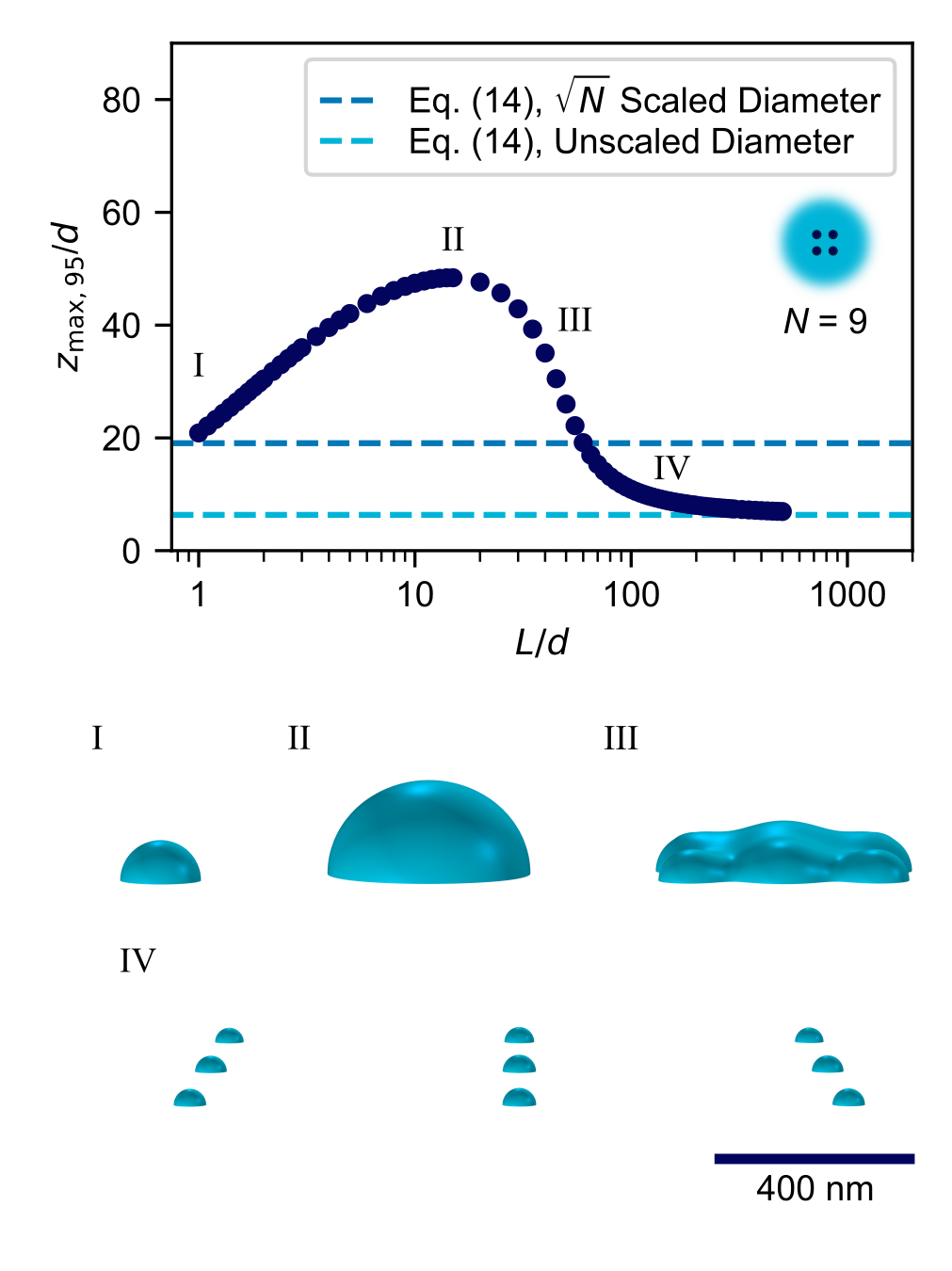}
\caption{\label{fig:07} The maximum vertical extent from the plane of the pores of the $\varphi = 0.95\frac{\Delta V}{2}$ equipotential surface, $z_{\mathrm{max},95}$ normalized by the diameter of the pores, $d$, as a function of the ratio of the interpore spacing to the pore diameter, $L/d$, for a square array of $N = 9$ pores surrounded by large hemispherical reservoirs. The dark blue dashed line indicates the vertical extent of the surface predicted by Eq.~\eqref{eq:singlePoreExtent} for a single pore with a larger effective diameter scaled by the square root of the number of pores, $d_\mathrm{eff} = \sqrt{N}d$, and the light blue dashed line indicates the vertical extent of the surface predicted by Eq.~\eqref{eq:singlePoreExtent} for a single pore where the diameter has not been scaled. The images below the graph show the equipotential surfaces for ($\mathrm{I}$) $L/d = 1$, ($\mathrm{II}$) $L/d=15$, ($\mathrm{III}$) $L/d=50$, and ($\mathrm{IV}$) $L/d=160$, which are plotted to scale. The corresponding positions of these equipotential surfaces are labeled on the graph.}
\end{figure}

In Fig.~\ref{fig:07}, we plot the values of $z_{\mathrm{max},95}$ extracted from the results of finite element calculations normalized by the diameter of a single pore, $d$, for various ratios of the interpore spacing to the pore diameter, $L/d$ for a square array of $N = 9$ pores surrounded by infinite hemispherical reservoirs. The value of $z_{\mathrm{max},95}$ grows to a maximum with increasing $L/d$ and then drops sharply to a value below what the value was initially at $L/d=1$. Below the graph, the equipotential surfaces for $L/d=1$, $L/d=15$, $L/d=50$, and $L/d=160$ are shown. A dark blue dashed line is located at the value of $z_{\mathrm{max},95}$ obtained from Eq.~\eqref{eq:singlePoreExtent} for a pore of effective diameter $d_\mathrm{eff} = \sqrt{N}d$, from which it can be seen that, as expected, for small values of the interpore spacing, $z_{\mathrm{max},95}$ approaches this value, confirming the behavior of the array as a single pore of larger effective diameter. Additionally, the plot of the equipotential surface for $L/d = 1$ located below the graph confirms that the shape of the equipotential surface is spheroidal, as would be expected for that of a single pore. A light blue dashed line is located at the value of $z_{\mathrm{max},95}$ obtained from Eq.~\eqref{eq:singlePoreExtent} for a single pore of unscaled diameter $d$. For large values of the interpore spacing, $z_{\mathrm{max},95}$ approaches this value. From the plot of the equipotential surface for $L/d = 160$ located below the graph, it is clear that the equipotential surface becomes localized around each pore and is roughly the shape that would be expected for isolated single pores of the same dimensions. When $z_{\mathrm{max},95}$ is maximum, the equipotential surface is still spheroidal, although it deforms rapidly and eventually becomes localized around each individual pore as the interpore spacing is increased. The value of $L/d$ for which $z_{\mathrm{max},95}$ reaches a maximum could be interpreted as the point at which, in terms of the access region, the array of pores stops behaving similarly to a single pore of increasing effective diameter and begins to transition towards behaving like that of independent pores. While this definition of coupling provides a satisfying physical description of the coupling, for complex pore arrangements, the exact geometry of the equipotential surfaces cannot be determined without first fully solving for the global potential profile, whereas the coupling length as we have defined it earlier can easily be obtained for linear and square arrays from Eq.~\eqref{eq:elllinear} and Eq.~\eqref{eq:ellsquare} respectively, or, for the case of arbitrarily arranged pores, derived directly from Eq.~\ref{eq:Greenwood} using only the system's geometric parameters.

\subsection{$N$ Pores, Cuboidal Reservoirs}

We now turn our attention to the case of finite size cuboidal reservoirs. Here, we consider the two array geometries separately. We also discuss the role that the length of the reservoirs has on the scaling of the conductance with pore number.

\subsubsection{Linear pore array}

\begin{figure*}[t]
\includegraphics[width=165.17mm]{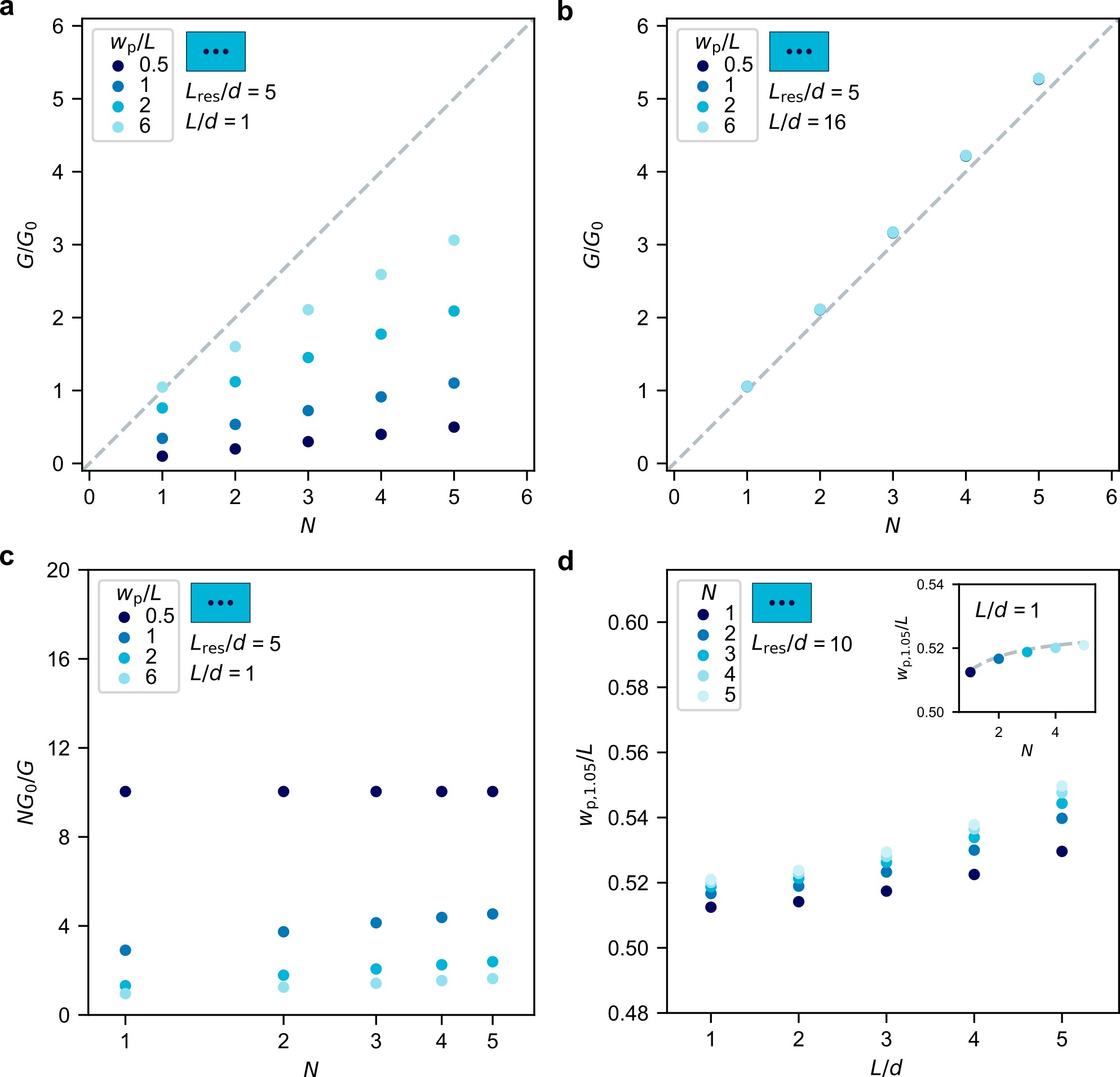}
\caption{\label{fig:08} The total conductance, $G$, normalized by $G_0$ for various ratios of the reservoir padding width, $w_\mathrm{p}$, to the interpore spacing, $L$, for different numbers of pores, $N$, in a linear array surrounded by cuboidal reservoirs of length $L_\mathrm{res} = 5d$ for (a) small, $L/d = 1$, and (b) large, $L/d = 16$, ratios of the interpore spacing to the nanopore diameter. (c) The number of pores, $N$, multiplied by $G_0$ and divided by the total conductance, $G$, is also plotted for different values of $N$ for the same geometries as in (a). Additionally, the padding width resulting in a $5\%$ increase in the total conductance from that of the commensurate case normalized by the interpore spacing, $w_{\mathrm{p},1.05}/L$, is plotted in (d) as a function of the ratio of the interpore spacing to the pore diameter, $L/d$, for different numbers of pores, $N$, in a linear array surrounded by cuboidal reservoirs of length $L_\mathrm{res}=10d$. The inset shows $w_{\mathrm{p},1.05}/L$ as a function of the number of pores, $N$. Dashed lines are located at $G/G_0 = N$ in (a) and (b). The dashed curve in the inset of (d) shows the results obtained by solving for $w_{\mathrm{p},1.05}$ in Eq.~\eqref{eq:linearBulkRatio}. The points in (b) nearly overlap.}
\end{figure*}

In Fig.~\ref{fig:08}(a) we plot the normalized conductance, $G/G_0$ for several values of $N$ for a linear array for padding width to interpore spacing ratios of $w_\mathrm{p}/L = 0.5,\ 1,\ 2,\ \mathrm{ and}\ 6$. Here, the interpore spacing is $L = d$ and the length of the reservoirs is $L_\mathrm{res}=5d$. A dashed line is plotted along with the data to indicate where linear $G = NG_0$ scaling falls on the graph. We see that for the incommensurate reservoir cases, the scaling is sublinear, whereas for the commensurate case (where $w_\mathrm{p}=L/2$), the scaling is linear. Also, the conductance overall decreases as $w_\mathrm{p}$ decreases, which reflects a transition between access resistance limited behavior at large $w_\mathrm{p}$ to bulk reservoir resistance limited behavior at small $w_\mathrm{p}$.

In Fig.~\ref{fig:08}(b) we plot the normalized conductance, $G/G_0$ for several values of $N$ for a linear array for padding width to interpore spacing ratios of $w_\mathrm{p}/L = 0.5,\ 1,\ 2,\ \mathrm{ and}\ 6$. Here, the interpore spacing is $L = 16d$ and the length of the reservoirs is $L_\mathrm{res}=5d$. A dashed line is plotted along with the data to indicate where linear $G = NG_0$ scaling falls on the graph. We see that, as expected, regardless of the padding width, the scaling relationship remains linear and falls onto the dashed line. Note that the slope of the data points here is greater than that of the dashed line, again due to the access resistance being lower than that given by Eq.~(\ref{eq:hallResistance}) since the reservoirs have a finite length.

To better resolve the sublinear scaling relationship, we plot $N(G/G_0)^{-1}$ versus the number of pores, $N$ on a logarithmic scale for a linear array with pore spacing $L=d$ and reservoir length $L_\mathrm{res}=5d$ for padding widths of $w_\mathrm{p} = L/2,\ L,\ 2L,\ \mathrm{and}\ 6L$ in Fig.~\ref{fig:08}(c). We see that for the commensurate case, the data points fall on a horizontal line located at approximately $2L_\mathrm{res}/d$, indicating linear $NG_0d/2L_\mathrm{res}$ scaling of the total conductance. For the case of large padding width, $w_\mathrm{p} = 6L$, we see a sloped line once again indicating $\sim N/\log{N}$ scaling of the conductance. The curves for the other padding widths fall in between these two cases. As $N$ increases, a transition from $\sim N/\log{N}$ scaling to $\sim N$ scaling will occur for these incommensurate cases as the bulk reservoir conductance (scales with $N$) eventually becomes the dominant factor. We discuss this in detail later for the case of square nanopore arrays.

Similar to the coupling length that we have defined for the case of infinite reservoirs, we can also define a padding width for which the conductance of an array approaches that of an array of the same pore arrangement but with a commensurate reservoir geometry. Specifically, we look at the value of the padding width for which the total conductance falls to within $5\%$ of the total conductance for the commensurate case, $w_{\mathrm{p},1.05}$. In Fig.~\ref{fig:08}(d) we plot $w_{\mathrm{p},1.05}$ normalized by the diameter of the pores, $d$, for various ratios of the interpore spacing to the pore diameter, $L/d$, for linear arrays consisting of $N$ = 1, 2, 3, 4, and 5 pores and a reservoir length of $L_\mathrm{res} = 10d$.

For very large values of $L/d$, the unit cells are dominated by the access resistance, which approaches its maximum value given by Eq.~\eqref{eq:hallResistance}. Consequently, increasing the padding width will not lead to significant increases in the conductance of the system, and so $w_{\mathrm{p},1.05}$ will tend towards infinity, reflecting the behavior seen in Fig.~\ref{fig:08}(b) where, regardless of the padding width, for large interpore spacings the values of the normalized conductance all collapse onto the same line. In Fig.~\ref{fig:08}(d) we consider when $L/d$ is very small, which is the more interesting case for which decreasing the padding width results in a transition in the scaling of the conductance with the number of pores from sublinear to linear, the contribution of the access resistance to the total resistance becomes negligible as $w_\mathrm{p}\rightarrow L/2$. Especially for the case most relevant to experiments  where  $L_\mathrm{res}$ is very large, the transition from  sublinear to linear scaling behavior is therefore dominated by changes in the bulk resistance of the system. As the padding width decreases, the cross sectional area of the system decreases as well, leading to an increase in the bulk resistance and bringing the total conductance down towards the value for a commensurate geometry.  For the commensurate case with a linear array of pores, the cross-sectional area is $NL^2$ while for the incommensurate case it is $4w_\mathrm{p}^2+2\left(N-1\right)w_\mathrm{p}L$. Since the transition to linear scaling is completely dominated by changes in the bulk resistance, the padding width at which the conductance falls to within $5\%$ of the conductance for the case of a commensurate geometry, $w_{\mathrm{p},1.05}$, can be easily found by solving
\begin{equation}
\label{eq:linearBulkRatio}
\frac{4w_{\mathrm{p},1.05}^2+2\left(N-1\right)w_{\mathrm{p},1.05}L}{NL^2} = 1.05.
\end{equation}
In the inset of Fig.~\ref{fig:08}(d) the value of $w_{\mathrm{p},1.05}$ normalized by $d$ is plotted for various values of $N$ for $L_\mathrm{res}/d = 10$ and $L/d = 1$. The gray dashed curve indicates the values obtained by solving Eq.~\eqref{eq:linearBulkRatio} for $w_{\mathrm{p},1.05}$, which agrees well with the results obtained from the finite element method calculations.

\subsubsection{Square pore array}

\begin{figure*}[th]
\includegraphics[width=165.17mm]{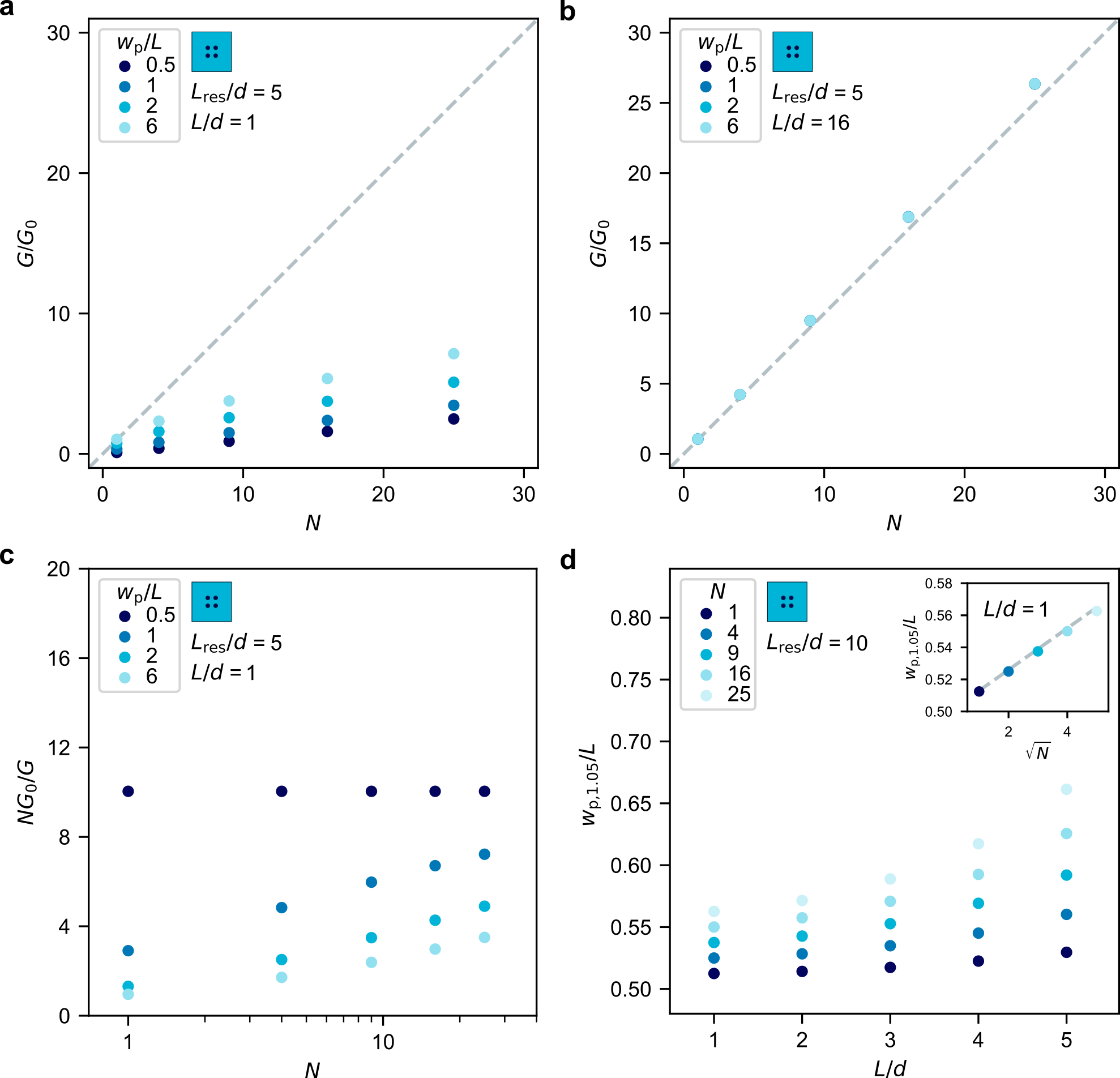}
\caption{\label{fig:09} The total conductance, $G$, normalized by $G_0$ for various ratios of the reservoir padding width, $w_\mathrm{p}$, to the interpore spacing, $L$, for different numbers of pores, $N$, in a square array surrounded by cuboidal reservoirs of length $L_\mathrm{res} = 5d$ for (a) small, $L/d = 1$, and (b) large, $L/d = 16$, ratios of the interpore spacing to the nanopore diameter. (c) The number of pores, $N$, multiplied by $G_0$ and divided by the total conductance, $G$, is also plotted for different values of $N$ for the same geometries as in (a). Additionally, the padding width resulting in a $5\%$ increase in the total conductance from that of the commensurate case  normalized by the interpore spacing, $w_{\mathrm{p},1.05}/L$, is plotted in (d) as a function of the ratio of the interpore spacing to the pore diameter, $L/d$, for different numbers of pores, $N$, in a square array surrounded by cuboidal reservoirs of length $L_\mathrm{res}=10d$. The inset shows $w_{\mathrm{p},1.05}/L$ as a function of the square root of the number of pores, $\sqrt{N}$. Dashed lines are located at $G/G_0 = N$ in (a) and (b). The dashed line in the inset of (d) shows the results obtained by solving for $w_{\mathrm{p},1.05}$ in Eq.~\eqref{eq:squareBulkRatio}. The points in (b) nearly overlap.}
\end{figure*}

In Fig.~\ref{fig:09}(a) we plot the normalized conductance, $G/G_0$ for several values of $N$ for a square array for padding width to interpore spacing ratios of $w_\mathrm{p}/L = 0.5,\ 1,\ 2,\ \mathrm{ and}\ 6$. The interpore spacing here is $L = d$ and the reservoir length is $L_\mathrm{res}=5d$. A dashed line is plotted along with the data to indicate where linear $G = NG_0$ scaling falls on the graph. For the incommensurate reservoir cases, the scaling is sublinear but approaches a linear scaling relationship for the commensurate case, where $w_\mathrm{p}=L/2$. For the case of linear scaling, the conductance is lower overall as, once again, the conductance transitions from being access resistance limited to bulk reservoir resistance limited.

In Fig.~\ref{fig:09}(b) we plot the normalized conductance, $G/G_0$ for several values of $N$ for a square array for padding width to interpore spacing ratios where the interpore spacing is $L = 16d$ and the length of the reservoirs is $L_\mathrm{res}=5d$. A dashed line is plotted along with the data to indicate where linear $G = NG_0$ scaling falls on the graph. Regardless of the padding width, the scaling relationship remains linear and falls approximately onto the dashed line, with the slope being slightly greater than 1 due to the finite length of the reservoirs as discussed previously for the case of linear arrays.

To better resolve the sublinear scaling relationship, we plot $N(G/G_0)^{-1}$ versus the number of pores, $N$, on a logarithmic scale for a square array with pore spacing $L=d$ and reservoir length $L_\mathrm{res}=5d$ for padding widths of $w_\mathrm{p} = L/2,\ L,\ 2L,\ \mathrm{and}\ 6L$ in Fig.~\ref{fig:09}(c). For the commensurate case, the data points fall on a horizontal line located at approximately $2L_\mathrm{res}/d$, indicating linear $NG_0d/2L_\mathrm{res}$ scaling of the total conductance. For the case of large padding width, $w_\mathrm{p} = 6L$, we see a curve shape indicating $\sim \sqrt{N}$ scaling of the conductance. The curves for the other padding widths fall in between these two cases.

In Fig.~\ref{fig:09}(d) we plot $w_{\mathrm{p},1.05}$ normalized by the diameter of the pores, $d$, for various ratios of the interpore spacing to the pore diameter, $L/d$, for square arrays consisting of $N$ = 1, 4, 9, 16, and 25 pores and a reservoir length of $L_\mathrm{res} = 10d$. As for the case of a linear array, for very large values of $L/d$, the unit cells are dominated by the access resistance and increasing the padding width will not lead to significant increases in the conductance of the system causing $w_{\mathrm{p},1.05}$ to tend towards infinity, as is apparent in Fig.~~\ref{fig:09}(b).

Fig.~\ref{fig:09}(d) shows the small $L/d$ range. As can be seen in the figure, $w_{\mathrm{p},1.05}/d$ increases both with $L/d$ and $N$. When $L/d$ is very small, the contribution of the access resistance to the total resistance becomes negligible as $w_\mathrm{p}\rightarrow 0.5L$ so that the transition from  sublinear to linear scaling behavior is controlled by changes in the bulk resistance of the system and therefore scales inversely with the cross sectional area of the reservoirs. For the commensurate case with a square array of pores, the cross-sectional area is $NL^2$ while for the incommensurate case it is $\left[\left(\sqrt{N}-1\right)L+2w_\mathrm{p}\right]^2$. As was true for the case of a linear array of pores, $w_{\mathrm{p},1.05}$, can be easily found by solving for the value from
\begin{equation}
\label{eq:squareBulkRatio}
\frac{\left[\left(\sqrt{N}-1\right)L+2w_{\mathrm{p},1.05}\right]^2}{NL^2} = 1.05.
\end{equation}
In the inset of Fig.~\ref{fig:09}(d), the value of $w_{\mathrm{p},1.05}$ normalized by $d$ is plotted for various values of $\sqrt{N}$ for $L_\mathrm{res}/d = 10$ and $L/d = 1$. The gray dashed line indicates the values obtained by solving Eq.~\eqref{eq:squareBulkRatio} for $w_{\mathrm{p},1.05}$, which agrees well with the results obtained from the finite element method calculations.

\begin{figure}[t]
\includegraphics[width=83.2mm]{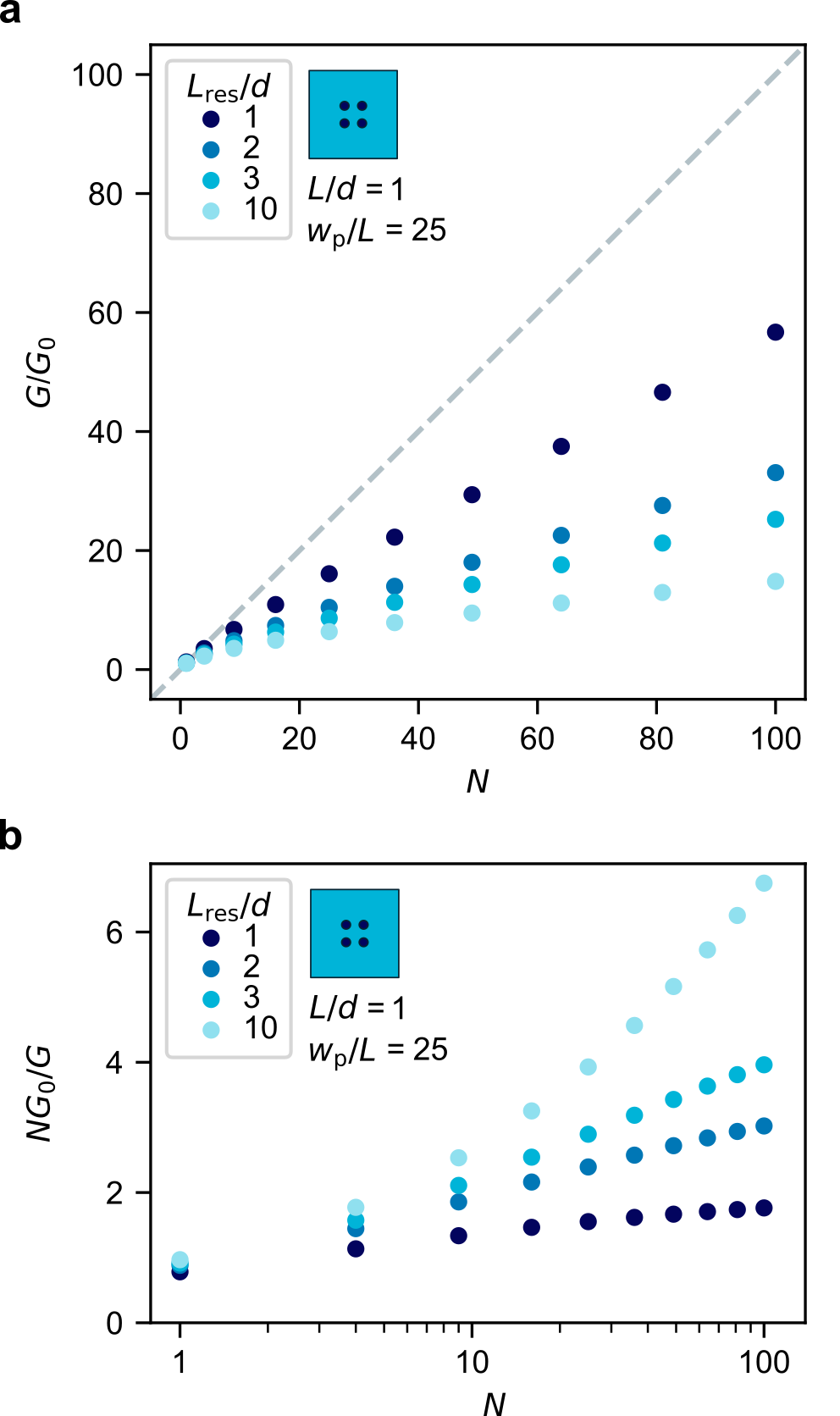}
\caption{\label{fig:10} (a) The total conductance, $G$, normalized by $G_0$ for various ratios of the length of the surrounding reservoirs, $L_\mathrm{res}$, to the diameter of the pores, $d$, for different numbers of pores, $N$, in a square array and (b) the number of pores, $N$, multiplied by $G_0$ and divided by the total conductance, $G$, for different numbers of pores, $N$, arranged in a square array and surrounded by cuboidal reservoirs for various ratios of the length of the surrounding reservoirs, $L_\mathrm{res}$, to the diameter of the pores, $d$. Here, the ratio of the interpore spacing to the pore diameter is $L/d=1$ and the ratio of the padding width to the interpore spacing is fixed at $w_\mathrm{p}/L = 25$.}
\end{figure}

We now turn our attention to the effects of changing the length of the reservoirs on the scaling of the conductance with pore number. Fig.~\ref{fig:10}(a) shows the normalized conductance, $G/G_0$, for several ratios of the reservoir length to the pore diameter, $L_\mathrm{res}/d$. Here, the interpore spacing is kept constant at $L=d$ and the padding width is kept constant at $w_\mathrm{p} = 25L$. Since the reservoirs are of finite size, there is a small contribution to the total resistance of the system from the bulk reservoir resistance. Although the padding width is kept constant, the total cross-sectional area of the system scales with $N$, as the entire width of the reservoirs increases as $\left(\sqrt{N}-1\right)L+2w_\mathrm{p}$. The bulk reservoir resistance term therefore decreases with $1/N$. 

Before examining how this bulk reservoir resistance term affects the scaling relationships for nanopore arrays surrounded by finite cuboidal reservoirs, we briefly comment on the effects of a bulk resistance decreasing with $1/N$ on the scaling relationships for infinite reservoirs in order for a comparison to be made. For the case of infinite reservoirs and closely spaced pores, the associated access resistance, as we have seen, decreases as $1/\sqrt{N}$. The bulk reservoir resistance is zero, and so the $1/\sqrt{N}$ scaling always holds. If the pores are given a finite thickness, so that each of the pores now also has a bulk pore resistance, the bulk resistance of the pores will scale as $1/N$ while the access resistance will still scale with $1/\sqrt{N}$. Since the bulk resistance drops more quickly with $N$, the access resistance will always dominate for large enough $N$, and so, regardless of the bulk resistance contribution of the pores, the $\sqrt{N}$ scaling of the conductance will eventually hold, as originally pointed out by Gadaleta et al.~\cite{Gadaleta2014}. This same phenomenon also applies to linear arrays, although the scaling of the access resistance is $\log{\left(N\right)}/N$ so that the conductance scales with $N/\log{N}$ for large enough $N$. However, for the case of finite reservoirs and closely spaced pores, the opposite trend, where the scaling of the conductance goes from sublinear to linear, can clearly be seen in Fig~\ref{fig:10}. While the bulk resistance still decreases with $1/N$, the access resistance now decreases much more rapidly than $1/\sqrt{N}$. This is because, for constant padding width and large $N$, the ratio of the width of the reservoirs to the width of the area occupied by the pores approaches unity so that, turning back to the qualitative description of closely spaced pores behaving similar to a single pore of larger effective diameter, the access resistance contribution drops to zero as predicted by Eq.~(\ref{eq:singalPoreModel}) and shown in Fig.~\ref{fig:03}. Since the length of the finite reservoir controls the bulk reservoir resistance contribution as well as how quickly the access resistance contribution approaches zero, the length of the reservoir sets the scale at which the transition from sublinear to linear scaling of the conductance occurs. Initially, the $\sim\sqrt{N}$ scaling dominates, but eventually, for large $N$, the linear component begins to dominate, and the scaling of the total conductance switches over to linear scaling with $N$. As can be seen in Fig.~\ref{fig:10}(a), this transition behavior of the scaling of the conductance does indeed occur and happens more quickly for reservoirs with smaller lengths. 

To better resolve this transition, in Fig.~\ref{fig:10}(b) we plot $N(G/G_0)^{-1}$ versus the number of pores, $N$, on a logarithmic scale for the same geometries that were plotted in Fig.~\ref{fig:10}(a). It is clear that for small values of $N$, the data points fall on a curve indicative of $\sim \sqrt{N}$ scaling, whereas for larger values of $N$ the value of $N(G/G_0)^{-1}$ begins to approach a horizontal line, indicative of linear scaling with $N$. For larger $L_\mathrm{res}$, this transition occurs for larger $N$, with the transition still being unobservable for $N=100$ in the case where $L_\mathrm{res}=10d$. 

\begin{figure}[t]
\includegraphics[width=83.2mm]{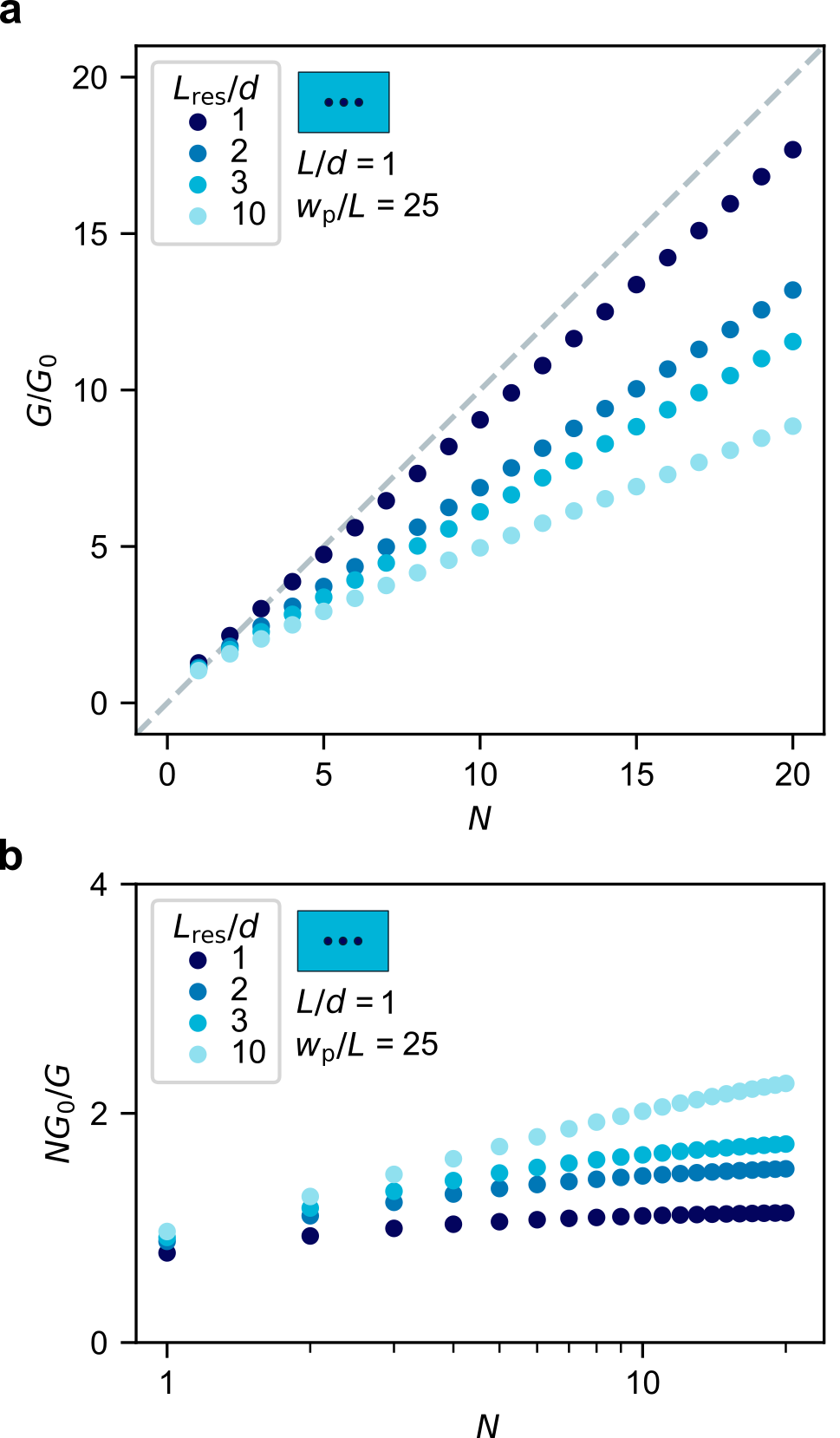}
\caption{\label{fig:11} (a) The total conductance, $G$, normalized by $G_0$ for various ratios of the length of the surrounding reservoirs, $L_\mathrm{res}$, to the diameter of the pores, $d$, for different numbers of pores, $N$, in a linear array and (b) the number of pores, $N$, multiplied by $G_0$ and divided by the total conductance, $G$, for different numbers of pores, $N$, arranged in a linear array and surrounded by cuboidal reservoirs for various ratios of the length of the surrounding reservoirs, $L_\mathrm{res}$, to the diameter of the pores, $d$. Here, the ratio of the interpore spacing to the pore diameter is $L/d=1$ and the ratio of the padding width to the interpore spacing is fixed at $w_\mathrm{p}/L = 25$.}
\end{figure}

As noted previously, a similar reservoir length dependent transition from sublinear to linear scaling also occurs for linear arrays of pores at large enough values of $N$. To show this, we plot the normalized conductance, $G/G_0$, for different ratios of the reservoir length to the pore diameter, $L_\mathrm{res}/d$, for several values of $N$ for the case of a linear array of pores in Fig.~\ref{fig:11}(a). Again, the interpore spacing is kept constant at $L=d$ and the padding width is kept constant at $w_\mathrm{p} = 25L$. In Fig.~\ref{fig:11}(b) we plot $N(G/G_0)^{-1}$ versus the number of pores, $N$, on a logarithmic scale for the same geometries that were plotted in Fig.~\ref{fig:11}(a). While it is difficult to distinguish between the sublinear and linear scaling behaviors in Fig.~\ref{fig:11}(a) it can clearly be seen in Fig.~\ref{fig:11}(b) that the initial $\sim N/\log{N}$ scaling transitions to linear scaling for large enough $N$. Again, increasing the the length of the reservoirs shifts the transition to larger values of $N$. 

Due to this transitioning behavior, it is therefore essential in these types of systems when predicting the scaling of conductance to consider both the contribution from the access resistance and the contribution from the bulk resistance. While here we consider only the case of negligible pore thickness, we note that for systems with pores of finite length, the bulk resistance of the nanopore can also play an important role. If the nanopore is long enough to be considered a very long channel, then the nanopore bulk resistance typically dominates for finite size reservoirs and a scaling linear with $N$ should be expected. A more thorough discussion of the effects of the finite length of the pores can be found in Appendix~D where we also derive a model based on Greenwood's approach for the conductance of arrays of finite length nanopores.

\section{Discussion and Conclusions}

We have found that, most generally, the scaling of conductance in nanopore arrays depends on two separate factors: the pore arrangement (set by array geometry, pore size, and interpore spacing) and the extent of the surrounding reservoirs. For large interpore spacing, the scaling of conductance with the number of pores is always linear. For incommensurate reservoirs, when the interpore spacing approaches the diameter of the pores, interactions between the pores cannot be ignored, and the conductance exhibits sublinear scaling with the number of pores. The exact form of the sublinear scaling depends on the arrangement of the pores, with $N/\log{N}$ scaling for linear arrays and $\sqrt{N}$ scaling for square arrays as previously reported~\cite{Gadaleta2014}, although, regardless of the arrangement, the absolute minimum scaling with the number of pores should be $\sqrt{N}$ scaling as the system approaches that of a single pore with larger effective diameter, $d_\mathrm{eff} = \sqrt{N}d$. The dependence here is well captured by the model derived by Gadaleta et al.~\cite{Gadaleta2014} and Greenwood~\cite{Greenwood1966}, and method of moments calculations (Appendix~B). As the reservoir geometry becomes commensurate, linear scaling of the conductance with pore number is recovered due to the condition of no flux across planes of symmetry as previously reported~\cite{Sebastian2023}, although for the case of very small interpore spacing the conductance is initially lower overall due to the dominance of the bulk resistance over the access resistance. Therefore, the dependence of the scaling of the conductance on the interpore spacing and the reservoir geometry fundamentally stems from their effects on the access resistance and the bulk resistance of the system and the strength of their contributions to the total resistance relative to each other.

An important practical consequence of the infinite-reservoir scaling is that the interpore spacing required to reach the linear, uncoupled regime is not fixed, but itself grows with the number of pores in the array. The characteristic coupling length, $\ell_\mathrm{c,95}$, was found to scale linearly with pore diameter $d$, but also with $\sim\log N$ for a linear array and $\sim\sqrt{N}$ for a square array. As a consequence, coupling can remain significant even when neighboring pores are separated by distances far exceeding their diameter: for example, for a $10\times10$ square array of pores with $d=1~\mathrm{nm}$, the nearest-neighbor spacing would need to exceed $\sim150~\mathrm{nm}$---more than two orders of magnitude larger than the pore diameter---for the array conductance to lie within 5\% of that of 100 independent pores. This illustrates that the common intuition that widely-spaced pores are necessarily uncoupled can break down substantially in large arrays, with direct implications for the design of nanopore array devices where independent-pore behavior is assumed.

For finite sized reservoirs, the maximum scaling of the conductance with pore number is actually greater than $NG_0$ for small reservoir lengths and large interpore spacing, due to the interplay of bulk reservoir resistance and decreased access resistance. In this regime, the system is dominated by the non-interacting access resistance of the pores, which is lower than the access resistance predicted for infinite reservoirs, leading to a larger conductance. Although linear, for the case of closely spaced pores, the smallest scaling of the conductance with the number of pores occurs for the case when the reservoir geometry is commensurate with the nanopore array. Here, the system behaves as the sum of individual single pore units, dominated by the bulk resistance rather than the access resistance, which for most relevant cases where $L_\mathrm{res}\gg d$, is much greater than the theoretical maximum value of the access resistance. Another interesting consequence of the finite length of the reservoirs is that, even for small interpore spacing where interactions between the pores cannot be ignored, the initial sublinear scaling of the conductance with pore number that is observed will eventually transition over to linear scaling, as long as the padding width is held constant. Turning back to the qualitative description of the closely spaced pore array as simply a pore with larger effective diameter, increasing the number of pores without increasing the padding width causes the ratio between the effective pore diameter and the width of the reservoir to approach 1, causing the total access resistance to go towards zero. The contribution from the access resistance therefore actually drops more quickly than $1/\sqrt{N}$ while the bulk resistance of the system still decreases with $1/N$, eventually leading to the dominance of the bulk term in the resistance and resulting in linear scaling of the conductance with $N$. This is in contrast to what happens for closely spaced pores and infinite reservoirs, as, regardless of any finite bulk resistance of the nanopores, the access resistance will eventually dominate for large numbers of pores since this bulk resistance will decrease with $1/N$ while the access resistance will always decrease less quickly as $1/\sqrt{N}$.

Since, for the commensurate case the system can be thought of as consisting of many identical single pore units, and, for the large padding width and small interpore spacing case the system can be thought of as a single pore with a larger effective pore size, it is useful to understand the case of a single pore with finite reservoirs on each side. For small padding widths, the conductance of the single pore system is clearly dominated by the bulk resistance of the reservoirs, while for large padding widths, the conductance is clearly dominated by the modified access resistance on each side of the pore, which is dependent on the length of the reservoirs and is lower than the access resistance predicted for infinite reservoirs as given by Eq.~(\ref{eq:hallResistance}). This is accurately captured by our model (Appendix~A) that improves upon existing models by using a more correct approximation for the current density in the pore which is essential for obtaining accurate results when the access resistance dominates. Our single pore model therefore allows to describe the conductance behavior of multipore systems in the two important limiting cases: 1) when the pores act as an array of single-pore unit cells, and 2) when the pores are closely packed together and effectively act as a single larger pore.

We also note that the condition of no flux across planes of symmetry must always hold. It has been pointed out, however, that while for finite reservoirs it may, in theory, always be possible to model the system as the sum of individual single pore units, for any cases other than the commensurate case, dividing the geometry into these individual units without first knowing the complete solution for the system may not be tractable in general, as these individual unit cells are no longer identical~\cite{Sebastian2023}. Therefore, for all but the simplest incommensurate cases, the unit cell description may be difficult to implement and it may be more informative to think of the scaling of the conductance in terms of where it falls relative to the maximum and minimum scaling relationships as we have done here. 

\begin{figure}[t]
\includegraphics[width=83.2mm]{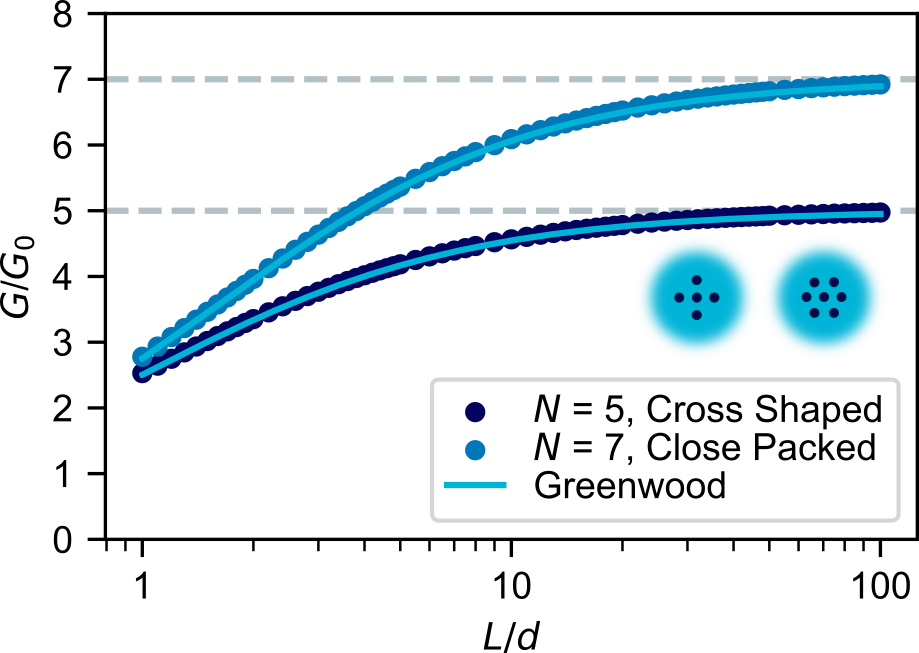}
\caption{\label{fig:12} The total conductance, $G$, normalized by $G_0$ for various ratios of the interpore spacing, $L$, to the nanopore diameter, $d$, for a five-pore-cross-shaped configuration and a seven-pore-close-packed configuration surrounded by large hemispherical reservoirs. The predictions of Eq.~(\ref{eq:Greenwood}) as derived by Greenwood~\cite{Greenwood1966} are plotted in the figure as solid blue curves. Gray dashed lines are shown at $G/G_0 = 5$ and $G/G_0=7$.}
\end{figure}

Although the specific sublinear scaling of the total conductance with $N$ for the case of large reservoirs and close pore spacing depends on the specific geometry of the pore array, the general trend of a transition from linear scaling at large values of the interpore spacing to sublinear scaling at low values of the interpore spacing with a lower bound at $\sqrt{N}G_0$ still holds. To demonstrate this, in Fig.~\ref{fig:12} we plot the normalized conductance $G/G_0$ as a function of the ratio of the interpore spacing to the pore diameter, $L/d$, for two different array geometries surrounded by infinite hemispherical reservoirs: a cross-shaped array geometry where $N=5$ and a close-packed geometry where $N=7$. It can be seen that for large $L/d$, the value of the normalized conductances for the two systems saturate at 5 and 7 respectively as indicated by the dashed lines. As expected, for small $L/d$, the value of the normalized conductance drops towards approximately $\sqrt{N}$. The predictions of Eq.~(\ref{eq:Greenwood}) are also plotted in Fig.~\ref{fig:12} for both the cross-shaped array and the close-packed array, once again demonstrating the close agreement between Eq.~(\ref{eq:Greenwood}) and the results of finite element calculations.

In conclusion, we have resolved the controversy surrounding the different reported scaling relationships for the conductance of nanopore arrays in the absence of surface charge. While previous models are able to accurately predict the scaling of the conductance of nanopore arrays with pore number, they are only able to do so when considering the appropriate reservoir geometry. None of these models can be universally applied to all systems. We show the transition between different scaling regimes and determine the relevant criteria for when each of these scaling relationships apply. The conductance scaling of nanopore array systems is highly dependent on both interpore spacing and reservoir geometry, and so both must always be taken into account in the effective design of nanofluidic systems. 

\begin{acknowledgments}
This work is based on a project supported by a Seed Research Award from the Brown University Initiative for Sustainable Energy (ISE). Z.G. acknowledges support through an Advanced Undergraduate Research Fellowship from the Brown University Summer/Semester Projects for Research, Internship, and Teaching (SPRINT) program.
\end{acknowledgments}

\section*{Author Declarations}

\subsection*{Conflict of Interest}

The authors have no conflicts to disclose.

\subsection*{Author Contributions}

\textbf{Jacob Bair:} Conceptualization (supporting); Data curation (lead); Formal Analysis (lead); Investigation (lead); Methodology (equal); Validation (lead); Visualization (lead); Writing – original draft (lead); Writing – review \& editing (equal).
\textbf{Zachery Gottshall:} Data curation (supporting); Formal Analysis (supporting); Investigation (supporting); Writing – original draft (supporting); Writing – review \& editing (equal).
\textbf{Matthias Kuehne:} Conceptualization (lead); Formal Analysis (supporting); Funding acquisition (lead); Investigation (supporting); Methodology (equal); Project Administration (lead); Resources (lead); Supervision (lead); Writing – original draft (supporting); Writing – review \& editing (equal).

\section*{Data Availability Statement}

The data that support the findings of this study are openly available in Zenodo at http://doi.org/10.5281/zenodo.21249209~\cite{dataset}.

\appendix

\section{Single pore analytical model}

We derive an analytical model for the electrical resistance of a single circular nanopore with radius $r_\mathrm{pore}=d/2$ located at the center of a rectangular membrane with side length $a$ and $b$ along the $x$- and $y$-direction, respectively. The cuboidal reservoirs on both sides of the pore have equal length $L_\mathrm{res}$. We consider the length of the pore itself to be negligible and the membrane to carry zero surface charge. We use a continuum-based approach and consider a sufficiently dilute electrolyte solution so that the convectionless transport of ions in our system is governed by the steady state Poisson-Nernst-Planck (PNP) equations consisting of Poisson's equation,
\begin{equation}
\label{eq:Poisson}
\nabla^2\varphi = -\frac{\rho_e}{\varepsilon}
\end{equation}
where $\varphi$ is the electric potential, $\varepsilon$ is the absolute permittivity of the electrolyte solution, and $\rho_e$ is electric charge density in the system, which is given by $\rho_e = e\left(c_+-c_-\right)$, where $e$ is the elementary charge and $c_+$ and $c_-$ are the positive and negative ion concentrations respectively, and the Nernst-Planck equations
\begin{equation}
\label{eq:NernstPlanck}\bm{j}_\pm = -D\left(\bm{\nabla} c_\pm \pm \frac{e}{k_BT}c_\pm\bm{\nabla}\varphi\right)
\end{equation}
where $\bm{j}_\pm$ is the flux of positive and negative ions, $k_B$ is Boltzmann's constant, $T$ is the temperature of the system, and $D$ is the diffusion coefficient, which we assume to be identical for the two ions. Since we are considering the steady state solution, $\bm{\nabla}\cdot\bm{j}_\pm=0$.
In the absence of surface charge, the concentration for both ion species is uniform throughout the system, $c_+ = c_- = c$, and so the PNP equations reduce to Laplace's equation for the electric potential, $\nabla^2\varphi = 0$. 

The general solution of Laplace's equation for a cuboidal geometry is given by
\begin{equation}
    \varphi(x,y,z)=A+Bz+\sum_m^\infty\sum_n^\infty a_{mn}\cos{(k_mx)}\cos{(k_ny)}\sinh{(\kappa_{mn}z)},
\end{equation}
with $k_m=m\pi/a$, $k_n=n\pi/b$, and $\kappa_{mn}=\sqrt{k_m^2+k_n^2}$, where $a$ and $b$ are the width and height of the cuboidal reservoir, as dictated by the no flux boundary condition at the walls of the system, $\bm{\nabla}\varphi\cdot\bm{n}=0$ where $\bm{n}$ is the normal vector to the walls, and the double sum excluding the case $m=n=0$.

At one end of the system, due to the applied voltage, we have $\varphi\left(x, y, 0\right) = \Delta V/2$ so that $A=\Delta V/2$. At the nanopore, which we define to exist in the plane located at $z = L_\mathrm{res}$, the current density through the pore is given by
\begin{equation}
i(x, y, L_\mathrm{res}) = -\sigma_\mathrm{bulk}\left(\frac{\partial\varphi}{\partial z}\right).
\end{equation}
In principle, the exact solution can be obtained without making any assumptions about the current density within the pore using only the condition of constant potential as required by the antisymmetry of the system, as has been done before for the similar problem of a constriction between two semi-infinite cylindrical conductors~\cite{Rosenfeld1981}. In order to obtain a more concise and easily evaluated form of the solution, we instead assume a radial inverse square root current density in the pore, i.e.,
\begin{equation}
\label{eq:isrcurrent}
    i(r)=\frac{I}{2\pi r_\mathrm{pore}^2\sqrt{1-r^2/r_\mathrm{pore}^2}},
\end{equation}
which is obtained by directly solving Laplace’s equation for the case of a circular constriction between two semi-infinite half-space conductors, and has been shown to be a good approximation for the actual distribution as long as the radius of the pore is small in comparison to the transverse dimensions of the surrounding reservoirs~\cite{Holm1967,Timsit1977,Rosenfeld1981,Bair2026}. Here, $I$ is the total current flowing through the system, and $r = \sqrt{x^2+y^2}$. We therefore have
\begin{equation}
\label{eq:derivativeCondition}
   \frac{\partial\varphi}{\partial z}\bigg |_{z=L_\mathrm{res}} = \begin{cases}
-\frac{i\left(r\right)}{\sigma_\mathrm{bulk}} & \text{for } 0\leq r < r_\mathrm{pore}, \\ 0 &\text{for } r \geq r_\mathrm{pore}.
\end{cases} 
\end{equation}

Multiplying both sides of Eq.~\eqref{eq:derivativeCondition} by $\cos{\left(k_px\right)}$ and $\cos{\left(k_qy\right)}$, we can utilize the orthogonality of the cosine functions by integrating over the proper domains to obtain
\begin{equation}
    B=-\rho_\mathrm{bulk}\frac{I}{4ab}
\end{equation}
and
\begin{equation}
    a_{mn}=-\frac{\rho_\mathrm{bulk}I\sin{(\kappa_{mn}r_\mathrm{pore})}}{abr_\mathrm{pore}\mathcal{N}_m\mathcal{N}_n\kappa_{mn}^2\cosh{(\kappa_{mn}L_\mathrm{res})}},
\end{equation}
where $\mathcal{N}_0=2$, and $\mathcal{N}_{i\ge1}=1$ with $i=m,n$.
Shifting the origin of the $z$-axis to the plane in which the pore lies and imposing $\varphi(x,y,0)=0$ by antisymmetry, we get
\begin{equation}
    R=R_\mathrm{bulk}+2R_\mathrm{a},
\end{equation}
with
\begin{equation}
    R_\mathrm{a}=\rho_\mathrm{bulk}\sum_{(m,n)\ne(0,0)} \frac{\sin{(\kappa_{mn}r_\mathrm{pore})\tanh{(\kappa_{mn}L_\mathrm{res})}}}{abr_\mathrm{pore}\mathcal{N}_m\mathcal{N}_n\kappa_{mn}^2}.
\end{equation}

If instead of Eq.~(\ref{eq:isrcurrent}) we assume constant current density in the pore, i.e.,
\begin{equation}
    i(r)=\frac{I}{\pi r_\mathrm{pore}^2},
\end{equation}
we get
\begin{equation}
\label{eq:greenRa}
    R_\mathrm{a}=\rho_\mathrm{bulk}\sum_{(m,n)\ne(0,0)} \frac{2J_1{(\kappa_{mn}r_\mathrm{pore})\tanh{(\kappa_{mn}L_\mathrm{res})}}}{abr_\mathrm{pore}\mathcal{N}_m\mathcal{N}_n\kappa_{mn}^2},
\end{equation}
where $J_1$ is the first order Bessel function. An expression similar to Eq.~(\ref{eq:greenRa}) has previously been derived by Sebastian and Green~\cite{Sebastian2023}.

\section{Method of moments}

To solve for the conductance of the nanopore array using the method of moments, we make use of the observation that the problem of resistance between electrodes in conducting media and the problem of capacitance between electrodes in insulating media are related by a simple transformation given by
\begin{equation}
\label{eq:proportionality}
    G = \frac{\sigma_\mathrm{bulk} C}{\varepsilon}
\end{equation}
where $G$ is the conductance, $C$ is the analogous capacitance, $\sigma_\mathrm{bulk}$ is the bulk conductivity of the conducting medium, and $\varepsilon$ is the permittivity of the insulating medium~\cite{Hall1975}. The analogous capacitance problem for our array of nanopores is an array of disks held at a constant potential. Following the same approach that has been used for single nanopores of different shapes~\cite{Liu2017}, we arbitrarily set this potential to be $\varphi=1~\mathrm{V}$, and solve for the charge density using the well documented method of moments~\cite{Gibson2021}. The integral equation for the electric potential at each point can be written as
\begin{equation}
    \varphi = 1~\mathrm{V} = \int_S\frac{\rho_\mathrm{e}\left(\bm{r}'\right)}{4\pi\varepsilon|\bm{r}-\bm{r}'|}\mathrm{d}S
\end{equation}
where $\rho_\mathrm{e}\left(\bm{r}'\right)$ is the electric charge density and $S$ is the total area of the disks. We divide each disk into a grid of $M$ squares of side length $2s$ and area $\Delta_\mathrm{s} = 4s^2$ and assume a constant charge density within each square. We choose $M$ independent observation and source points at the center of each square to serve as our $\bm{r}$ and $\bm{r}'$. We then have the matrix equation
\begin{equation}
    \left[Z\right]\left[\rho_\mathrm{e}\right] = 1
\end{equation}
where the elements of $Z$ are given by
\begin{equation}
    Z_{mn}=\frac{\Delta_s}{4\pi\varepsilon\sqrt{\left(x_m-x_n\right)^2+\left(y_m-y_n\right)^2}}
\end{equation}
for $m\neq n$ and
\begin{equation}
    Z_{mm}=\frac{2s}{\pi\varepsilon}\log{\left(1+\sqrt{2}\right)}
\end{equation}
where $m$ and $n$ index the entire set of squares in the system and therefore run from 1 to $N\cdot M$, where $N$ is the number of disks. From this matrix equation, we are able to solve for the assumed constant charge density in each of the squares.

We then proceed to calculate the capacitance of the nanopore array by taking the sum of the individual contributions to the total charge and dividing by the electric potential. Using Eq.~(\ref{eq:proportionality}), we arrive at the conductance of the  nanopore array system.

\section{Coupling length expressions}

From Eqs.~(\ref{eq:hallConductance}) and (\ref{eq:Greenwood}), the conductance of $N$ coupled pores relative to the conductance of $N$ independent pores, $G_0^{(N)} = N\sigma_\mathrm{bulk}d$, is
\begin{equation}
    \frac{G}{G_0^{(N)}} = \left(1+\frac{d}{\pi N}\sum_{i\neq j}\frac{1}{L_{ij}}\right)^{-1}.
\end{equation}
Setting this ratio equal to $p$ (e.g.\ $p=0.95$) and solving for the double sum gives the defining condition for $\ell_\mathrm{c}(p,N)$, independent of array geometry:
\begin{equation}
    \sum_{i\neq j}\frac{1}{L_{ij}} = \frac{2\pi N(1-p)}{2p\,d}
\end{equation}
which we can alternatively write as
\begin{equation}
\label{eq:mastercond}
    S(N,\ell) = \frac12\sum_{i\neq j}\frac{1}{L_{ij}} = \frac{\pi N(1-p)}{2pd}.
\end{equation}

For pores on a line with spacing $\ell$, $L_{ij}=|i-j|\,\ell$. Grouping unordered pairs $(i<j)$ by separation $k=j-i=1,\dots,N-1$, there are $(N-k)$ pairs at each $k$, so
\begin{align}
    S(N,\ell) &= \frac{1}{\ell}\sum_{k=1}^{N-1}\frac{N-k}{k}
    = \frac{1}{\ell}\left(N\sum_{k=1}^{N-1}\frac1k - (N-1)\right) \\
    &= \frac{N H_{N-1} - (N-1)}{\ell}
\end{align}
where $H_n=\sum_{k=1}^nk^{-1}$ is the harmonic number. Substituting into Eq.~(\ref{eq:mastercond}) at $\ell=\ell_\mathrm{c,linear}$ and solving for $\ell_\mathrm{c,linear}$:
\begin{equation}
    \ell_\mathrm{c,linear}(p,N) = \frac{2pd}{\pi(1-p)}\left(H_{N-1}+N^{-1}-1\right).
\end{equation}
Using the large-$N$ asymptotic form $H_{N-1}\approx \log N + \gamma$ (with $\gamma\approx0.5772$ the Euler constant) and $1/N\to0$ yields the $\log N$ scaling:
\begin{equation}
    \ell_\mathrm{c,linear}(p,N)\approx\frac{2pd}{\pi(1-p)}\left(\log N+\gamma-1\right).
\end{equation}

For an $s\times s$ array ($N=s^2$) with spacing $\ell$, pores sit at $(i,j)\ell$, $i,j=0,\dots,s-1$. In the continuum limit ($s\to\infty$), the discrete double sum is approximated by an integral over a square domain of side $\Lambda=s\ell$ with site density $\ell^{-2}$:
\begin{equation}
    S(N,\ell) \approx \frac{1}{2\ell^2}\int_0^{\Lambda}\!\!\int_0^{\Lambda}\!\!\int_0^{\Lambda}\!\!\int_0^{\Lambda}
    \frac{\mathrm{d}x_1\,\mathrm{d}y_1\,\mathrm{d}x_2\,\mathrm{d}y_2}{\sqrt{(x_1-x_2)^2+(y_1-y_2)^2}}.
\end{equation}
Substituting $u=x_1-x_2$, $v=y_1-y_2$ collapses the $(x_1,x_2)$ and $(y_1,y_2)$ integrations into overlap-weighted 1D integrals of width $(\Lambda-|u|)$ and $(\Lambda-|v|)$ respectively:
\begin{align}
    \int_0^{\Lambda}\!\!\int_0^{\Lambda}\!\!\int_0^{\Lambda}\!\!\int_0^{\Lambda}\frac{\mathrm{d}x_1\,\mathrm{d}y_1\,\mathrm{d}x_2\,\mathrm{d}y_2}{\sqrt{(x_1-x_2)^2+(y_1-y_2)^2}} \\
     = \int_{-\Lambda}^{\Lambda}\!\!\int_{-\Lambda}^{\Lambda} \frac{(\Lambda-|u|)(\Lambda-|v|)}{\sqrt{u^2+v^2}}\,\mathrm{d}u\,\mathrm{d}v.
\end{align}
Rescaling $u=\Lambda p'$, $v=\Lambda q'$ and using the even symmetry in $u,v$ to restrict to the first quadrant:
\begin{align}
\int_{-\Lambda}^{\Lambda}\!\!\int_{-\Lambda}^{\Lambda} \frac{(\Lambda-|u|)(\Lambda-|v|)}{\sqrt{u^2+v^2}}\,\mathrm{d}u\,\mathrm{d}v \\
    = 4\Lambda^3\int_0^1\!\!\int_0^1\frac{(1-p')(1-q')}{\sqrt{p'^2+q'^2}}\,\mathrm{d}p'\,\mathrm{d}q' = 4\Lambda^3 K,
\end{align}
where
\begin{equation}
    K = \frac{1-\sqrt2}{3}+\log(1+\sqrt2)\approx0.74330.
\end{equation}
With $\Lambda=s\ell$, this gives $S(N,\ell)\approx \dfrac{2K}{\ell}s^3 = \dfrac{2K}{\ell}N^{3/2}$. Substituting into Eq.~(\ref{eq:mastercond}) at $\ell=\ell_\mathrm{c,square}$:
\begin{equation}
    \ell_\mathrm{c,square}(p,N)\approx\frac{4Kpd}{\pi(1-p)}\sqrt{N},
\end{equation}
revealing the $\sqrt N$ scaling.

\section{Finite length pores}
While we focus on the ideal case of pores that are negligibly thin, for many real systems, including for the case of pores in 2D materials when their diameters are relatively small, the length of the pores approaches or exceeds their diameter. The scaling behaviors that we observe for negligibly thin pores capture the effects of overlapping access resistance regions as well as the effects of bulk reservoir resistance for the case of finite cuboidal reservoirs. For pores where the access resistance still dominates over the bulk resistance of the pores, which corresponds to when $l_\mathrm{n} \ll \frac{\pi d}{4}$ for pores of length $l_\mathrm{n}$ and diameter $d$, the length of the pores can still be considered negligible and the scaling relationships that we have described are expected to hold. When the bulk resistance of the pores completely dominates over the access resistance, which corresponds to the case where $l_\mathrm{n} \gg \frac{\pi d}{4}$, the conductance will typically scale linearly with $N$ since the bulk resistance of the pores simply adds in parallel provided that the no flux boundary condition remains valid at the walls of the pores, although for cases approaching that of infinite reservoirs, as discussed in Sec. III D 2, sublinear scaling for closely spaced pores may be recovered for large enough $N$. For intermediate cases, both the access resistance and the bulk resistance of the pores must be considered, although qualitatively the scaling behavior will remain similar to the case of negligibly thin pores with only minor modifications to account for their finite length. For example, calculations of the unit cell conductance would need to be modified to include the bulk resistance of the finite length pore which simply adds in series to the access resistance terms. The correct baseline to compare the conductance of the arrays to would also become the conductance of a single pore of finite length connecting two semi-infinite electrolyte reservoirs, given by
\begin{equation}
G_{0}^{*} = \sigma_\mathrm{bulk}\left(\frac{1}{d}+\frac{4l_\mathrm{n}}{\pi d^2}\right)^{-1}
\end{equation}
rather than simply $G_0$ as given by Eq.~\eqref{eq:conductivity}, which only accounts for the access resistance terms on each side of the pore. Additionally, depending on the length of the pores, the access resistance terms may be up to $5\%$ greater for pores of finite length than for negligibly thick pores, which follows from applying the Maxwell-Rayleigh correction from electrical contact theory for the spreading resistance of a long cylindrical conductor to finite length pores, which accounts for the change in the current density at the mouth of the pores due to their finite length~\cite{Holm1967}.

\begin{figure*}[t]
\includegraphics[width=165.17mm]{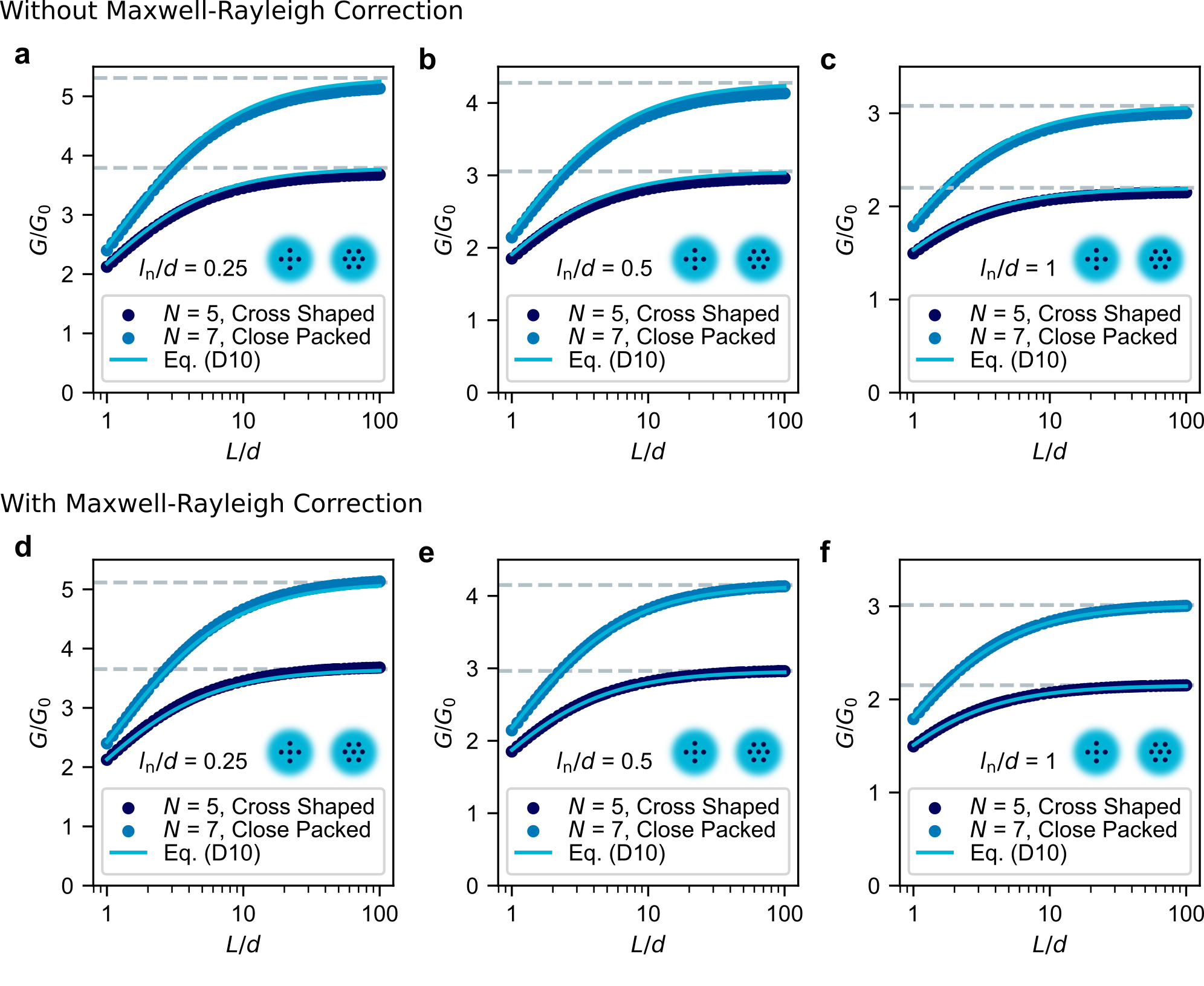}
\caption{\label{fig:13} The total conductance, $G$, normalized by $G_0$ for various ratios of the interpore spacing, $L$, to the nanopore diameter, $d$, for a five-pore-cross-shaped configuration and a seven-pore-close-packed configuration surrounded by large hemispherical reservoirs for pores of finite length, $l_\mathrm{n}$. The ratios of the nanopore length to the diameter are $l_\mathrm{n}/d = 0.25$ for (a) and (d), $l_\mathrm{n}/d = 0.5$ for (b) and (e), and $l_\mathrm{n}/d = 1.0$ for (c) and (f). The predictions of Eq.~\eqref{eq:finiteLength} are plotted as solid blue curves where the Maxwell-Rayleigh correction has not been applied to the self-interaction terms for (a), (b) and (c) and has been applied for (d), (e), and (f). Gray dashed lines are shown at $G = 5G_0^*$ and $G=7G_0^*$, where again the Maxwell-Rayleigh correction has not been applied for (a), (b) and (c) and has been applied for (d), (e), and (f).}
\end{figure*}

Specifically for the case of infinite reservoirs, by applying the monopole approximation, it is relatively straightforward to obtain an analytical expression for the total conductance of the system. We first consider $N$ circular nanopores of radii $a_i$ and negligible length at planar positions $\mathbf r_i$ at the interface between two infinite electrolyte reservoirs of bulk resistivity $\rho_\mathrm{bulk}$. Following Greenwood's approach for solving the related electrical contact spot problem \cite{Greenwood1966}, the potential at nanopore $i$ due to the currents $I_j$ flowing through all of the nanopores in the system is
\begin{equation}
\varphi_i = \sum_j b_{ij} I_j ,
\qquad
b_{ii} = \frac{\rho_\mathrm{bulk}}{4a_i}, \qquad
b_{ij} = \frac{\rho_\mathrm{bulk}}{2\pi s_{ij}} \ \ (i\neq j),
\label{eq:bij}
\end{equation}
where $s_{ij} = |\mathbf r_i - \mathbf r_j|$ and the term $b_{ij}$ is the leading monopole approximation to the interaction term between nanopores $i$ and $j$. Since all nanopores lie on the same equipotential interface,
\begin{equation}
\varphi_i = \frac{\Delta V}{2} \qquad \text{for all } i,
\label{eq:equipotential}
\end{equation}
where $\Delta V$ is the potential difference applied across the entire system. We now consider the case when the nanopores are given a small but finite length, $l_\mathrm{n}$. To extend the derivation based on Greenwood's approach, we redefine $\varphi_i$ to be the potential in each nanopore at the equipotential surface running through the midplanes of the pores. We add the bulk resistance of each finite length pore to the leading approximation for the self-interaction term for that pore so that we now have
\begin{equation}
b_{ii} = \frac{\rho_\mathrm{bulk}}{4}\left(\frac{1}{a_i}+\frac{2 l_\mathrm{n}}{\pi a_i^2}\right)
\end{equation}
and
\begin{equation}
    \varphi_i
 =\frac{\rho_\mathrm{bulk}}{4}\left(\frac{I_i}{a_i}+\frac{2I_il_\mathrm{n}}{\pi a_i^2}+\frac{2}{\pi}\sum_{j\neq i}
\frac{I_j}{s_{ij}}\right) \qquad \text{for all } i.
 \end{equation}
Assuming, as Greenwood did, that the current through each individual nanopore is inversely proportional to the self-interaction term for that pore, 
\begin{equation}
I_i = \lambda \frac{1}{b_{ii}},
\end{equation}
where $\lambda$ is the proportionality constant, then the total current through the system can be found as 
\begin{equation}
I = \lambda\sum_i\frac{1}{b_{ii}},
\end{equation}
and the rate of heat production, or equivalently, the power dissipated, is given by
\begin{equation}
    W = I\Delta V = 2\sum_i\varphi_iI_i = 2\lambda^2\sum_i\sum_j\frac{b_{ij}}{b_{ii}b_{jj}}.
\end{equation}
The total resistance of the system is then given by
\begin{equation}
\begin{split}
    R = \frac{W}{I^2} = &\frac{\rho_\mathrm{bulk}}{2}\left(\sum\limits_{i}\frac{\pi a_i^2}{\pi a_i+2l_\mathrm{n}}\right)^{-1} \\
    &+\frac{\rho_\mathrm{bulk}}{\pi}\left(\mathop{\sum\sum}_{i \neq j}\frac{\pi^2a_i^2a_j^2}{s_{ij}\left(\pi a_i+2l_\mathrm{n}\right)\left(\pi a_j+2l_\mathrm{n}\right)}\right) \\
    &\times \left(\sum_i\frac{\pi a_i^2}{\pi a_i+2l_\mathrm{n}}\right)^{-2},
\end{split}
\end{equation}
and the total conductance of the system can be found as usual from $G = R^{-1}$.
If all of the pores are of the same radius, $a$, then the expression can further be simplified as 
\begin{equation}
\label{eq:finiteLength}
\begin{split}
    R = &\frac{\rho_\mathrm{bulk}}{2}\left(\sum\limits_{i}\frac{\pi a^2}{\pi a+2l_\mathrm{n}}\right)^{-1} \\
    &+\frac{\rho_\mathrm{bulk}}{\pi}\left(\mathop{\sum\sum}_{i \neq j}\frac{1}{s_{ij}\left(\pi a+2l_\mathrm{n}\right)^2}\right) \\
    &\times \left(\sum_i\frac{1}{\pi a+2l_\mathrm{n}}\right)^{-2}.
\end{split}
\end{equation}
For the simplest case of two identical nanopores of the same diameter, $d=2a$, and length, $l_\mathrm{n}$, with a center-to-center distance, $L$, the resistance is given by
\begin{equation}
    R = \frac{1}{2}\rho_\mathrm{bulk} \left(\frac{1}{d}+\frac{4l_\mathrm{n}}{\pi d^2}+\frac{1}{ \pi L}\right),
\end{equation}
and the conductance can be written as
\begin{equation}
    G = 2\sigma_\mathrm{bulk}\left(\frac{1}{d}+\frac{4l_\mathrm{n}}{\pi d^2}+\frac{1}{ \pi L}\right)^{-1}.
\end{equation}

To validate the results of our analytical model and to demonstrate that, in general, arrays of pores of finite length will behave in a way consistent with what we have described, we performed finite element method calculations to obtain the total conductance for cross-shaped and close-packed arrays of identical finite length pores surrounded by large hemispherical reservoirs. The results of these calculations along with the predictions of Eq.~\eqref{eq:finiteLength} are plotted in Fig.~\ref{fig:13} for several ratios of the length of the nanopores to the diameter of the nanopores, $l_\mathrm{n}/d$, as a function of the ratio of the center-to-center distance between the pores to the diameter of the pores, $L/d$. The expression we derived agrees well with the numerically calculated results. As expected, the general transition from sublinear scaling of the conductance with $N$ at small values of $L/d$ to linear scaling of the conductance with $N$ at large values of $L/d$ still occurs. The interpore spacing at which the conductance appears to saturate also decreases with increasing pore length, indicating a transition towards the linear scaling behavior of systems dominated by the bulk resistance of the pores. The value to which the conductance saturates for large $L/d$ is also $NG_0^*$ as expected, which is indicated by the gray dashed lines in Fig.~\ref{fig:13}. It should also be noted that Eq.~\eqref{eq:finiteLength} provides a slight overestimate of the conductance in all cases for the finite length pores as shown in Fig.~\ref{fig:13}(a), (b), and (c). If the $5\%$ Maxwell-Rayleigh correction is applied to the access resistance terms for both $G_0^*$ and the self-interaction terms in our derivation so that
\begin{equation}
G_{0}^{*} = \sigma_\mathrm{bulk}\left(\frac{1.05}{d}+\frac{4l_n}{\pi d^2}\right)^{-1}
\end{equation}
and
\begin{equation}
b_{ii} = \frac{\rho_\mathrm{bulk}}{4}\left(\frac{1.05}{a_i}+\frac{2 l_\mathrm{n}}{\pi a_i^2}\right),
\end{equation}
then the agreement of the analytical model and the value to which the conductance should saturate to with the results of the finite element calculations improves for $l_\mathrm{n}/d = 0.5$ and $l_\mathrm{n}/d = 1.0$ as shown in Fig.~\ref{fig:13}(e) and (f), although the conductance is now underestimated for $l_\mathrm{n}/d = 0.25$ as shown in Fig.~\ref{fig:13}(d), as the pore is likely not yet long enough for the full correction term to accurately be applied.

\bibliography{porearrays}

\end{document}